\documentclass[aps,showpacs,eqsecnum,twocolumn,superscriptaddress,usegraphicx]{revtex4}

\usepackage{amsmath}
\usepackage{latexsym}
\usepackage{graphicx}
\usepackage{psfig}

\begin{document}


\title{{\tt NADA}: A new code for studying self-gravitating tori around black holes}


\author{Pedro J. Montero}
\affiliation{Departamento de Astronom\'{\i}a y Astrof\'{\i}sica,
Universidad de Valencia, Dr. Moliner 50, 46100 Burjassot, Spain}

\author{Jos\'{e} A. Font}
\affiliation{Departamento de Astronom\'{\i}a y Astrof\'{\i}sica,
Universidad de Valencia, Dr. Moliner 50, 46100 Burjassot, Spain}

\author{Masaru Shibata}
\affiliation{Graduate School of Arts and Sciences, University of
  Tokyo, Komaba, Meguro, Tokyo 153-8902, Japan}

\date{\today}

\begin{abstract}
We  present a  new two-dimensional numerical code called {\tt Nada}
designed to solve the full Einstein equations coupled to the general
relativistic hydrodynamics equations. The code is mainly intended for
studies of self-gravitating accretion disks (or tori) around black
holes, although it is also suitable for regular spacetimes. Concerning
technical aspects the Einstein equations are formulated and solved in
the code using a formulation of the standard 3+1 (ADM) system, the
so-called BSSN approach. A key feature of the code is that derivative
terms in the spacetime evolution equations are computed using a
fourth-order centered finite difference approximation in conjunction
with the Cartoon method to impose the axisymmetry condition under
Cartesian coordinates (the choice in {\tt Nada}), and the
puncture/moving puncture approach to carry out black hole
evolutions. Correspondingly, the general relativistic hydrodynamics
equations are  written in flux-conservative form and solved with
high-resolution,  shock-capturing schemes. We perform and discuss a
number of tests to assess the accuracy and expected convergence of the
code, namely (single) black hole evolutions, shock tubes, and
evolutions of both spherical and rotating relativistic stars in equilibrium,
the gravitational collapse of a spherical relativistic star leading to the
formation of a black hole. In addition, paving the way for specific applications of the code, we also present results from fully general relativistic numerical simulations of a system formed by a black hole surrounded by a self-gravitating torus in equilibrium. 
\end{abstract}


\pacs{
04.25.Dm, 
04.40.Dg, 
04.70.Bw, 
95.30.Lz, 
97.60.Jd
}


\maketitle

\section{Introduction}
\label{sec:introduction}

Self-gravitating tori (or thick accretion disks) orbiting black holes (BHs) are common end-products in a number of scenarios of relativistic astrophysics. Theoretical evolutionary paths predict that they may form after the merger of a binary system formed by a BH and a neutron star (NS) or the system formed by two NS (see e.g.\cite{Baiotti08a} and references therein). In addition, they may also be the result of the gravitational collapse of the rotating core of massive stars \cite{Woosley93,Paczynski98}. State-of-the-art numerical simulations have started to provide quantitative estimates of the viability of such systems to form \cite{Sekiguchi04,Sekiguchi07,Stephens07,Shibata06b,Shibata99d,Shibata03c,Marronetti04,Shibata05,Shibata06a,Baiotti08a,Shibata06,Shibata07b}. Furthermore, such a thick disk plus BH system is thought to be the central engine for gamma-ray bursts \cite{Piran99,Aloy20,Aloy05,Meszaros06}. Therefore, understanding the formation and dynamics of such systems is a highly relevant enterprise, along with discerning its stability properties. Namely, whether they may be subject to axisymmetric and non-axisymmetric instabilities which could also lead to the emission of a significant amount of gravitational radiation.

In particular, the so-called runaway instability, first found by
Abramowicz, Calvani and Nobili \cite{Abramovici83}, is an axisymmetric
instability that could destroy the torus on  dynamical timescales. The
numerical study of the runaway instability in general relativity has
so far been investigated under different assumptions and
approximations (see e.g. \cite{Font02a,Daigne04} and references
therein). Despite some progress has been made the existing works are
not still conclusive on the likelihood of the instability, mainly due to the absence of important physics in the modeling. The complexity of handling the presence of a spacetime singularity in addition to the (magneto-)hydrodynamics and the self-gravity of the (possibly magnetized) accretion torus, turn very challenging the task of carrying out full general relativistic simulations of BH surrounded by a self-gravitating torus. The code we present in this paper is built with this mid-term goal in mind.

Numerical studies of the dynamics of matter around BHs are abundant in
the literature (see e.g.~the references reported
in~\cite{Font03}). For the kind of specific work we discuss in this
paper we note, in particular, that \cite{Brandt97e,Brandt98} already
focused on the numerical evolution of matter in dynamical axisymmetric
BH spacetimes. Recent developments in numerical relativity associated
with the puncture method for evolving BHs \cite{Campanelli06,Baker06}
have proved essential to perform accurate and long-term stable
evolutions of spacetimes containing BHs, and the first successful
simulations of  BH/BH binaries have only been possible very recently
(see~\cite{Pretorius07} and references therein). Shibata
\cite{Shibata06,Shibata07b} used the puncture method to investigate
the merger of BH/NS binary system in full general relativity, and
Faber et al.~\cite{Faber07} considered relativistic spherical accretion onto a BH. In addition, Baiotti and Rezzolla \cite{Baiotti06} were also able to handle the accretion of matter onto a newly formed BH resulting from the collapse of a rotating NS without needing to excise the singularity.

In order to take advantage of this approach, Shibata \cite{Shibata07}
presented a formulation for computing equilibrium configurations of a
BH and a self-gravitating torus in the puncture framework. Previous
computations of this system by Nishida and Eriguchi \cite{Nishida94},
and also by Ansorg and Petroff \cite{Ansorg05} were not done in the
puncture framework. In this paper we use the initial data in the
formalism presented in \cite{Shibata07} to carry out the first
dynamical simulations of a self-gravitating torus in equilibrium
around a BH as a first step towards a systematic investigation of the
runaway instability in full general relativity. Such initial data are
evolved using a new two-dimensional, general relativistic
hydrodynamics code called {\tt Nada} which is presented and thoroughly
tested in this paper. As we explain in detail below our code solves
the 3+1 Einstein equations using the formulation originally proposed
by Nakamura~\cite{Nakamura87} and subsequently modified by
Shibata-Nakamura~\cite{Shibata95} and
Baumgarte-Shapiro~\cite{Baumgarte99}, which is usually known as the
BSSN formulation. On the other hand, the general relativistic
hydrodynamics equations are written in flux-conservative form and
solved with a high-resolution shock-capturing scheme
(HRSC)~\cite{Banyuls97,Font03,Shibata03}. In addition, the {\tt Nada}
code implements the Cartoon method \cite{Alcubierre01b} which allows
to impose the axisymmetry condition while still admitting the use of
Cartesian coordinates, the puncture/moving puncture approach to
deal with the BH singularity and derivative terms in the spacetime
evolution equations are computed using a fourth-order centered finite 
difference approximation. 

The code is written in FORTRAN 90, requires approximately $8$ bytes
of memory per grid point, and it takes about $200$ microseconds per
grid point per time step using a five stages Runge-Kutta (see section
III) for the time integration on a 2.0 GHz AMD Opteron Dual Core 270.

The organization of the paper is as follows: the formulation of the Einstein equations, including the implementation of the puncture approach and gauge conditions, along with the formulation of the general relativistic equations is briefly presented in
Section II. Section III gives a short description of the boundary
conditions and numerical methods employed for the time
evolutions. Section IV and V are devoted to present results from
tests the code has passed, for vacuum and non-vacuum spacetimes,
respectively. Section VI discusses numerical simulations of
self-gravitating equilibrium tori, in preparation for subsequent work.
A summary of our conclusions is given in Section VII. We use units in
which $c=G=1$, and in sections V. C to V. F we also use $M_{\odot}=1$. Greek indices run from 0 to 3, Latin indices from 1 to 3, and we adopt the standard convention for the summation over repeated indices. 

\section{Basic equations}
\label{equations}

We give next a brief overview of the formulation for the system of
Einstein and hydrodynamic equations as have been implemented in the code.

\subsection{Formulation of Einstein equations}   
\label{feqs}

\subsubsection{BSSN formulation}

        We follow the 3+1 formulation in which the spacetime is foliated into a set of non-intersecting spacelike hypersurfaces. In
this approach, the line element is written in the following form
\begin{equation}
ds^2 = -(\alpha^{2} -\beta _{i}\beta ^{i}) dt^2 + 
        2 \beta_{i} dx^{i} dt +\gamma_{ij} dx^{i} dx^{j}, 
\end{equation}
where $\alpha$, $\beta^i$ and $\gamma_{ij}$ are the lapse function,  the shift three-vector, and the three-metric, respectively. The latter is defined by 
\begin{equation}
\gamma_{\mu\nu}=g_{\mu\nu}+n_{\mu}n_{\nu},
\end{equation}
where $n^{\mu}$ is a timelike unit-normal vector orthogonal to a spacelike hypersurface.

A reformulation of the ADM system, the BSSN formulation~\cite{Nakamura87,Shibata95,Baumgarte99}, has been implemented to solve the Einstein equations. Initially, a conformal factor is introduced, and the conformally related metric is written as
\begin{equation}
\label{cmetric}
{\tilde{\gamma}}_{ij}=e^{-4\phi}\gamma_{ij},
\end{equation} 
such that the determinant of the conformal metric,
$\tilde{\gamma}_{ij}$, is unity and $\phi=\ln(\gamma)/12$, where $\gamma=\det(\gamma_{ij})$. We also define the conformally related traceless part of the extrinsic curvature $K_{ij}$, 
\begin{equation}
\label{caij}
 {\tilde{A}}_{ij}=e^{-4\phi}A_{ij}=e^{-4\phi}\left(K_{ij}-\frac{1}{3}\gamma_{ij}K\right),
\end{equation}
where $K$ is the trace of the extrinsic curvature. After introducing these new variables, the evolution equations can be expressed as
\begin{align}
\left(\partial_{t}- {\cal L}_{\beta}\right)\phi & = -\frac{1}{6}\alpha K,
\label{ecfactor} \\
\left(\partial_{t}- {\cal L}_{\beta}\right) K & = -\gamma^{ij}D_{j}D_{i}\alpha+
	\alpha \tilde{A}_{ij}\tilde{A}^{ij}\nonumber \\ +
	& \frac{1}{3}\alpha K^{2} + 4\pi\alpha\left(E+S\right),
\label{ek} \\
\left(\partial_{t}- {\cal L}_{\beta}\right)\tilde{\gamma}_{ij} & = -2\alpha \tilde{A}_{ij},
\label{ecmetric}\\
\left(\partial_{t}- {\cal L}_{\beta}\right)\tilde{A}_{ij} & = 
	e^{-4\phi}\left(-D_{i}D_{j}\alpha+
	\alpha\left(R_{ij}-8\pi S_{ij}\right)\right)^{\mathcal {TF}} \nonumber \\
      & +\alpha\left(K\tilde{A}_{ij}-2\tilde{A}_{il}\tilde{A}^{l}_{j}\right),
\label{eaij}\\
\partial_{t}\tilde{\Gamma}^{i} & =-2\tilde{A}^{ij}\partial_{j}\alpha+2\alpha\left(\tilde{\Gamma}^{i}_{jk}\tilde{A}^{jk}-\frac{2}{3}\tilde{\gamma}^{ij}\partial_{j}K\right. \nonumber \\
&\left.
 -8\pi\tilde{\gamma}^{ij}S_{j}+6\tilde{A}^{ij}\partial_{j}\phi\right)+ \beta^{j}\partial_{j}\tilde{\Gamma}^{i}-\tilde{\Gamma}^{j}\partial_{j}\beta^{i} \nonumber \\
&	+\frac{2}{3}\tilde{\Gamma}^{i}\partial_{j}\beta^{j}+\frac{1}{3}\tilde{\gamma}^{li}\partial_{jl}\beta^{j}+\tilde{\gamma}^{lj}\partial_{lj}\beta^{i},
\label{eCCF}
\end{align} 
where  $\tilde{\Gamma}^{i}_{\hskip 0.2cm jk}$ are the connection
coefficients associated with $\tilde{\gamma}_{ij}$, ${\cal L}_{\beta}$
refers to the Lie derivative along the shift vector (see
e.g. \cite{Alcubierre02a} for the tensor weights necessary to evaluate
the Lie derivatives), and $D_{i}$ and $R_{ij}$ are the covariant derivative operator and the three-dimensional Ricci tensor associated with the three-metric $\gamma_{ij}$, respectively. The object $\tilde{\Gamma}^{i}$, known as the {\it{conformal connection functions}}, is defined as
\begin{equation}
\label{CCF}
\tilde{\Gamma}^{i}\equiv \tilde{\gamma}^{jk}\tilde{\Gamma}^{i}_{\hskip
0.2cm jk}=-\partial_{j}\tilde{\gamma}^{ij}.
\end{equation}

The Ricci tensor $R_{ij}$ that appears in the source term of the
evolution equation (\ref{eaij}) is split into two parts as follows
\begin{equation}
\label{riccisplit}
R_{ij}=R^{\phi}_{ij} + \tilde{R}_{ij},
\end{equation}
where $R^{\phi}_{ij}$ is given by
\begin{align}
\label{ricciphi}
R^{\phi}_{ij} & =-2\tilde{D}_{i}\tilde{D}_{j}\phi-2\tilde{\gamma}_{ij}\tilde{D}^{k}\tilde{D}_{k}\phi\nonumber
\\ & +4\tilde{D}_{i}\phi\tilde{D}_{j}\phi-4\tilde{\gamma}_{ij}\tilde{D}^{k}\phi\tilde{D}_{k}\phi,
\end{align}
where $\tilde{D}_{i}$ is the covariant derivative with respect to the
conformal metric $\tilde{\gamma}_{ij}$. The conformal Ricci tensor,
$\tilde{R}_{ij}$, is expressed as
\begin{align}
\label{riccic}
\tilde{R}_{ij} &
=-\frac{1}{2}\tilde{\gamma}^{mn}\partial_{mn}\tilde{\gamma}_{ij}+\tilde{\gamma}_{k(i}\partial_{j)}\tilde{\Gamma}^{k}+\tilde{\Gamma}^{k}\tilde{\Gamma}_{(ij)k}\nonumber
\\
& +\tilde{\gamma}^{mn}\left(2\tilde{\Gamma}^{k}_{m(i}\tilde{\Gamma}_{j)kn}+\tilde{\Gamma}^{k}_{in}\tilde{\Gamma}_{kmj}\right).
\end{align}

During the evolution we also enforce the constraints
${\rm Tr} (\tilde{A}_{ij})=0$  and ${\rm det} (\tilde{\gamma}_{ij})=1$ at
every time step by using the following substitutions
\begin{equation}
\label{Acons}
\tilde{A}_{ij} \rightarrow \tilde{A}_{ij}-\frac{{\rm Tr} (\tilde{A}_{ij})}{3}\tilde{\gamma}_{ij},
\end{equation}
\begin{equation}
\label{Gcons}
\tilde{\gamma}_{ij} \rightarrow \tilde{\gamma}_{ij}/\tilde{\gamma}^{1/3}.
\end{equation}
The matter source terms, $E$, $S_{i}$, and $S_{ij}$ appearing in the Einstein equations are 
projections of the stress-energy tensor $T^{\mu\nu}$ on the
hypersurface with respect to the unit normal $n^{\mu}$
\begin{eqnarray}
E&=&n_{\mu}n_{\nu}T^{\mu\nu}, \nonumber \\
S_{i}&=&-\gamma_{i\mu}n_{\nu}T^{\mu\nu}, \nonumber \\
S_{ij}&=&\gamma_{i\mu}\gamma_{j\nu}T^{\mu\nu}, 
\label{ecmatsourc}
\end{eqnarray}
with $S=S_{ij} \gamma^{ij}$.

In addition to the evolution equations there are three constraint
equations, the Hamiltonian, the momentum and the Gamma constraints, which are only used as diagnostics of the accuracy of the numerical evolutions

\begin{align}
\label{ham1}
  \mathcal{H} & \equiv \tilde{\gamma}^{ij}\tilde{D}_{i}\tilde{D}_{j}e^{\phi}
  -\frac{e^{\phi}}{8}\tilde{R}+\frac{e^{5\phi}}{8}\tilde{A}_{ij}\tilde{A}^{ij}\nonumber \\
  & -\frac{e^{5\phi}}{12}K^{2}+2\pi e^{5\phi}E = 0, \\
\label{mom1}
  \mathcal{M}^i &\equiv \tilde{D}_j(e^{6\phi}\tilde{A}^{ij}) -
  \frac{2}{3}e^{6\phi}\tilde{D}^{i}K-8\pi e^{6\phi}\tilde{\gamma}^{ij}S_{j} = 0,\\
\label{Gam1}
  \mathcal{G}^i &\equiv \tilde{\Gamma}^{i}+\partial_{j}\tilde{\gamma}^{ij}=0.
\end{align}

\subsubsection{Puncture approach}

Recent breakthroughs in numerical relativity have finally made possible accurate and long-term stable 3+1 evolutions of singular spacetimes, including the challenging cases of the collision of compact binaries formed by either two BHs or a BH and a NS (see~\cite{Pretorius07} and references therein). One of the key ingredients for such success has been the so-called ``puncture approach''~\cite{Brandt97b}, in which BH initial data are modeled by the Brill-Lindquist topology \cite{Brill63}, where  a ``throat'' at the BH horizon connects two asymptotically flat regions. One of the asymptotically flat ends is compactified to a single point known as the puncture, leading to a coordinate singularity.

The puncture approach has the advantage that its numerical
implementation within the BSSN formalism is rather simple. The
original proposal only considered the  {\it fixed} puncture approach
\cite{Bruegmann97,Alcubierre00b,Alcubierre02a},  where the conformal
factor is split into a regular part and a singular part. Although this
method does not lead to long-term stable evolutions of spacetimes
containing BHs, it allows for a number of code tests. However, two
different groups \cite{Campanelli06,Baker06} developed recently the
{\it moving} puncture approach in which no singular term of the
conformal factor is factored out, and the punctures are allowed to move through the grid. The difference between these two methods is how the conformal factor is evolved. In the so-called $\phi$-method \cite{Baker06}, the original BSSN variable $\phi$ is evolved through the usual BSSN evolution Eq.~(\ref{ecfactor}). On the other hand, the so-called $\chi$-method \cite{Campanelli06} introduces a new conformal factor defined as $\chi\equiv e^{-4\phi}$, and the following evolution equation, that replaces Eq.~(\ref{ecfactor}),
\begin{equation}
\label{chi}
\left(\partial_{t}- {\cal L}_{\beta}\right)\chi = \frac{2}{3}\chi (\alpha K-\partial_{j}\beta^{j}).
\end{equation}
This moving puncture approach, together with the so-called puncture
gauge (see Sec. II A.3), has led to  major success in simulations of
binary BHs
\cite{Campanelli07,Baker07,Gonzalez07,Marronetti08,Rezzolla08,Pollney07,Herrmann07}.
Motivated by this, we have implemented both methods in our 2D code,
the $\phi$-method and the $\chi$-method, for the moving puncture
approach to investigate the dynamics of matter around BHs.
%
\subsubsection{Gauge choices}
%
In addition to the BSSN spacetime variables (${\tilde{\gamma}_{ij}},{\tilde{A}_{ij}},K,\phi$ or $\chi,\tilde{\Gamma}^{i}$), there are two more  variables left undetermined, the  lapse, $\alpha$ and the shift vector, $\beta^{i}$. The code can handle arbitrary gauge conditions, however long-term BH evolutions in the moving puncture framework have been successful with some combination of the so called {``1+log''} condition \cite{Bona97a} for the lapse, and the {``Gamma-freezing''} condition for the shift vector \cite{Alcubierre02a}. 

The form of this slicing condition most commonly used for moving puncture evolutions contains the advective term $\beta^{i}\partial_{i}\alpha$ and is expressed as
\begin{equation}
\label{1+log0}  
\partial_{t}\alpha - \beta^{i}\partial_{i}\alpha = -2 \alpha K. 
\end{equation}
Several authors (see, e.g.~\cite{Hannam07b,Bruegmann06,Faber07}) have used the so-called {``non-advective 1+log''}, by dropping   the advective term in Eq.~(\ref{1+log0}), for single BH evolutions. In this case, the slicing condition takes the form
\begin{equation}
\label{1+log1}  
\partial_{t}\alpha = -2 \alpha K. 
\end{equation}
It was shown by \cite{Hannam07b}, that the evolution of a single
puncture using the condition given by Eq.~(\ref{1+log1}) settles down
into a time-independent and maximally sliced solution. 

For the simulations of a BH surrounded by a self-gravitating torus presented in this paper, we choose the slicing given by Eq.~(\ref{1+log1}) as the BH does not move through the grid, and therefore the advective term $\beta^{i}\partial_{i}\alpha$ does not play an important role.

For the shift vector, we choose the Gamma-freezing condition, written as 
\begin{equation}
\label{shift1}  
\partial_{t}\beta^{i} = \frac{3}{4}B^{i},
\end{equation}
\begin{equation}
\label{shift2}  
 \partial_{t}B^{i} = \partial_{t}\tilde{\Gamma}^{i}-\eta B^i,
\end{equation}
where $\eta$ is a constant that acts as a damping term, introduced
both to prevent long term drift of the metric functions and to prevent
oscillations of the shift vector \cite{Alcubierre02a}. This parameter
also has an effect on the coordinates of the final spacelike
hypersurface. In Ref.~\cite{Bruegmann06} it was shown  that increasing the value of
$\eta$ increases the coordinate size of the BH, which allows for
better numerical resolution across the BH. On the other hand, larger
values of the damping parameter introduce a higher drift in the
location of the horizon in time, as well as in the deformation of the
metric during the evolution, caused by the larger values reached by
the connection functions $\tilde{\Gamma}^{i}$ \cite{Meter06,Faber07}. Bearing this in mind,
we use $\eta=0.3/M$, where $M$ is the ADM mass of the system, for the evolutions of spacetimes containing a BH 
presented in this paper.

\subsection{Formulation of the hydrodynamics equations}    
 The general relativistic hydrodynamics equations, expressed through the conservation  equations for the stress-energy tensor $T^{\mu\nu}$ and the continuity equation are
\begin{equation}
\label{hydro eqs}
\nabla_\mu T^{\mu\nu} = 0\;,\;\;\;\;\;\;
\nabla_\mu \left(\rho u^{\mu}\right) = 0,
\end{equation}
where $\rho$ is the rest-mass density, $u^{\mu}$ is the fluid
four-velocity and $\nabla$ is the covariant derivative with respect to
the spacetime metric. Following \cite{Shibata03} the general relativistic hydrodynamic equations are written in a
conservative form in cylindrical coordinates. Since the
Einstein equations are solved only in the $y=0$ plane with Cartesian
coordinates, the hydrodynamic equations are rewritten in the Cartesian coordinates for $y=0$. 
The following definitions for the hydrodynamical variables are used
\begin{equation}
\label{def1}
\rho_{*}\equiv \rho W e^{6\phi},
\end{equation}
\begin{equation}
v^{i}\equiv \frac{u^{i}}{u^{t}}=-\beta^{i}+\alpha\gamma^{ij}\frac{\mbox{\^{u}}_{j}}{hW},
\end{equation}
\begin{equation}
\mbox{\^{u}}_{i}\equiv h u_{i},
\end{equation}
\begin{equation}
\mbox{\^{e}}\equiv
\frac{e^{6\phi}}{\rho_{*}}T_{\mu\nu}n^{\mu}n^{\nu}=hW-\frac{P}{\rho W},
\end{equation}
\begin{equation}
\label{def2}
W\equiv \alpha u^{t},
\end{equation}
where $W$ and $h$ are the Lorentz factor and the specific fluid
enthalpy respectively, and $P$ is the pressure.

By defining the vector of unknowns, ${\bf{U}}$, and fluxes ${\bf{F}}^{x}$
and ${\bf{F}}^{z}$ along the $x$ and $z$ directions as
\begin{equation}
{\bf{U}}=(\rho_{*},J_{x},J_{y},J_{z},E_{*}),
\end{equation}
\begin{eqnarray}
{\bf{F}}^{x}=[\rho_{*}v^{x},J_{x}v^{x}+P\alpha
\sqrt{\gamma},J_{y}v^{x},J_{z}v^{x},\nonumber \\
E_{*}v^{x}+P\alpha \sqrt{\gamma}(v^{x}+\beta^{x})],
\end{eqnarray}
\begin{eqnarray}
{\bf{F}}^{z}=[\rho_{*}v^{z},J_{x}v^{z},J_{y}v^{z},J_{z}v^{z}+P\alpha
\sqrt{\gamma},\nonumber \\
E_{*}v^{z}+P\alpha \sqrt{\gamma}(v^{z}+\beta^{z})],
\end{eqnarray}
with  $J_{i}\equiv\rho_{*}\mbox{\^{u}}_{i}$ and $E_{*}\equiv
\rho_{*}\mbox{\^{e}}$, the set of hydrodynamic equations (\ref{hydro eqs})
can be written in conservative form as 
\begin{equation}
\label{cylcon}
\partial_{t}{\bf{U}}+\partial_{x}{\bf{F}}^{x}+\partial_{z}{\bf{F}}^{z}={\bf{S}},
\end{equation}
where ${\bf{S}}$ is the vector of sources. We refer to \cite{Shibata03}
for further details on these equations, in particular regarding the form of
the source terms.

To close the system of equations, we choose two possible equations of state, the so-called $\Gamma$-law equation of state (ideal fluid) given by 
\begin{equation}
\label{EOS1}
P=\left(\Gamma -1\right)\rho\epsilon ,
\end{equation}
where $\epsilon$ is the specific internal energy, 
and a polytropic equation of state
\begin{equation}
\label{EOS2}
 P={\kappa} \rho^{\Gamma}.
\end{equation}
Here $\kappa$ is the polytropic constant, $\Gamma=1+1/N$ and $N$ is the polytropic index. In those simulations where the system evolves adiabatically, such that no shocks are present, we use the polytropic equation of state during the evolution.
 
After each time iteration the conserved variables
 (i.e.~$\rho_{*},J_{x},J_{y},J_{z},E_{*}$) are
updated and the {\it primitive} hydrodynamical variables (i.e.~$\rho,v^{x},v^{y},v^{z},\epsilon$) have to be
recovered at the corresponding step. This is done by solving the  following
equation for the Lorentz factor, $W$, derived from the
normalization of the 4-velocity of the fluid
\begin{equation}
\label{con2prim}
W^{2}=1+\gamma^{ij}u_{i}u_{j}= 1+
\gamma^{ij}\mbox{\^{u}}_{i}\mbox{\^{u}}_{j}\left(\frac{\mbox{\^{e}}}{W}+\frac{P}{\rho
W^{2}}\right)^{-2}.
\end{equation}
Once solved for $W$, the other variables, $\rho,v^{i},P,\epsilon$
and $h$ are computed from Eqs.~(\ref{def1})-(\ref{def2}) and the equation of state.

\section{Numerics}
\label{sec:numerical-methods}

There are two schemes implemented for the update of the numerical
solution with time: an iterative Crank-Nicholson (ICN) scheme taking
two corrector steps~\cite{Teukolsky00}, and an optimal strong
stability-preserving (SSP) Runge-Kutta of fourth-order algorithm with
5 stages~\cite{Spiteri} (RK4). The RK4 scheme is used for simulations of spacetimes
containing a singularity, where a high-order of accuracy is an
important issue.

 We use second-order slope limiter reconstruction schemes (both minmod
 and MC are implemented in the code) to obtain the left and right states of the primitive variables (i.e. $\rho,v^{x},v^{y},v^{z},\epsilon$) at each cell interface,  and  these reconstructed variables are then used to compute the left and right states of the evolved quantities $(\rho_{*},J_{x},J_{y},J_{z},E_{*})$. Next, we use HLLE or  Roe approximate solvers to compute the numerical fluxes in the $x$ and $z$ directions. Details on such high-resolution shock-capturing schemes are available elsewhere (see e.g.~\cite{Font03} and references therein).

Derivative terms in the spacetime evolution equations are represented by second or fourth order centered finite difference approximation \cite{Zlochower05} in a uniform Cartesian grid except for the advection terms (terms formally like $\beta^{i}\partial_{i}u$), for which an upwind scheme is used.

\subsection{Discretization of Axisymmetric Systems: \\
``Cartoon'' Method}

 As pointed out by \cite{Alcubierre01b} any given system of equations in 3D possessing a rotation symmetry with respect to the $z$-axis can be finite differenced and  solved in the $x-z$ $(y=0)$ plane alone because of the symmetry condition. When the system of equations is expressed in Cartesian coordinates, partial derivative terms with respect to the $y$-coordinate will appear and these need to be computed using the computational domain. For instance, in a second-order centered difference approximation, the values of the derivative in the $y$-direction of a quantity $f(x,0,z)$, are computed using the values of $f$ at the nearest two grid points {\it i.e.} $f(x,-\Delta y,z)$, $f(x,\Delta y,z)$. Then, because of axisymmetry, the $y$-derivatives can be determined in the $x-z$ plane from information contained in this same plane. The Cartoon method obtains the boundary conditions at $y=\pm \Delta y$ that are necessary to evaluate the derivatives in the $y$-direction by means of a rotation about the symmetry axis of the different tensor quantities. On the other hand, in a fourth-order centered difference approximation, the values of the derivative in the $y$-direction of a quantity $f(x,0,z)$, are computed using the values of $f$ at the grid points $f(x,\pm 2\Delta y,z)$ and $f(x,\pm \Delta y,z)$. 

The values of the variables at the positions $(\sqrt{x^{2}+{\Delta
    y}^{2}},0,z)$ and $(\sqrt{x^{2}+{4\Delta y}^{2}},0,z)$ are
interpolated from the neighboring grid points using Lagrange
polynomial interpolation (see e.g.~\cite{Press92}) with interpolating polynomials of degree 2 and 4, depending on the order of the finite difference approximation. 

\subsection{Boundary Conditions}
\label{sec:BCs}

The computational domain is defined as $0\leq x \leq L$ and $0 \leq z
\leq L $, where $L$ refers to the location of the outer
boundaries. For simulations with the puncture method we used a
staggered Cartesian grid to avoid that the location of the puncture at
the origin coincides with a grid point.
A number of different boundary conditions are implemented in the
code. These are imposed for the spacetime variables or the
hydrodynamical primitive variables at the inner and outer boundaries as follows: for both the spacetime and hydrodynamical variables, $\pi$-rotation symmetry is imposed around the $z$-axis, and equatorial plane symmetry with respect to the $z=0$ plane. At the outer boundaries we  impose radiative boundary conditions \cite{Alcubierre02a}. Note that we do not apply this boundary condition to the conformal connection functions, $\tilde{\Gamma}^{i}$, for which we used static boundary conditions.

\subsection{Atmosphere treatment}
\label{sec:atm-treatment}

An important ingredient in numerical simulations based on finite
difference schemes to solve the hydrodynamic equations is the
treatment of vacuum regions. The standard approach is to add an
atmosphere of very low density filling these regions
\cite{Font02c}. We follow this approach and treat the atmosphere as a
perfect fluid with a rest-mass density several orders of magnitude
smaller than that of the bulk matter. The hydrodynamic equations are
solved in the atmosphere region in the same way that is done for the
region of the bulk matter. If the conservative variables $\rho_{*}$ or
$E_{*}$ fall below some minimum value then the values of the conserved
quantities are set to the atmosphere value. Similarly in the routine
that converts primitive variables from conservatives, if the rest-mass
density $\rho$ or specific internal energy $\epsilon$ fall below the
value set for the atmosphere, this point is reset to have the
atmosphere value of the primitive variables. In particular for
simulations of relativistic stars, and systems composed of a BH plus a 
self-gravitating torus system, the atmosphere density is usually taken
to be about $6$-$8$ orders of magnitude smaller than the initial
maximum rest-mass density.

\subsection{Diagnostics}
\label{sec:Diag}
To check the accuracy of the numerical simulations we monitor the
violation of the Hamiltonian constraint, and the conservation of the
total rest-mass $M_{*}$, the ADM mass $M$, and the angular momentum $J$. We compute these
various quantities in the $y=0$ plane as 

\begin{equation}
\label{Mass}
M_{*}=4\pi\int_{0}^{L}{x}dx\int_{0}^{L}{\rho_{*}}dz,
\end{equation}
\begin{eqnarray}
\label{MassADM}
M=-2\int_{0}^{L}{x}dx\int_{0}^{L}{}dz\left[-2\pi E
e^{5\phi}+\frac{e^{\phi}}{8}\tilde{R}\right. \nonumber \\ 
\left.-\frac{e^{5\phi}}{8}\left(\tilde{A}_{ij}\tilde{A}^{ij}-\frac{2}{3}K^{2}\right)\right],
\end{eqnarray}
\begin{equation}
\label{ang}
J=4\pi\int_{0}^{L}{x^{2}}dx\int_{0}^{L}{\rho_{*}\tilde{u}_{y}}dz.
\end{equation}

In axisymmetry the apparent horizon (AH hereafter) equation becomes a nonlinear
ordinary differential equation (ODE) for the AH shape function,
$h=h(\theta)$~\cite{Shibata97a,Thornburg2007}. We have implemented an AH
finder that solves this ODE by a shooting method using that
$\partial_{\theta}h(\theta=0)=0$ and
$\partial_{\theta}h(\theta=\pi/2)=0$ as boundary conditions. We define
the mass of the AH as 
\begin{equation}
\label{Mah}
M_{\rm {AH}}=\sqrt{\frac{\mathcal{A}}{16\pi}},
\end{equation}
where $\mathcal{A}$ is the area of the AH.

\begin{figure*}
\includegraphics[width=8.7cm, angle=0]{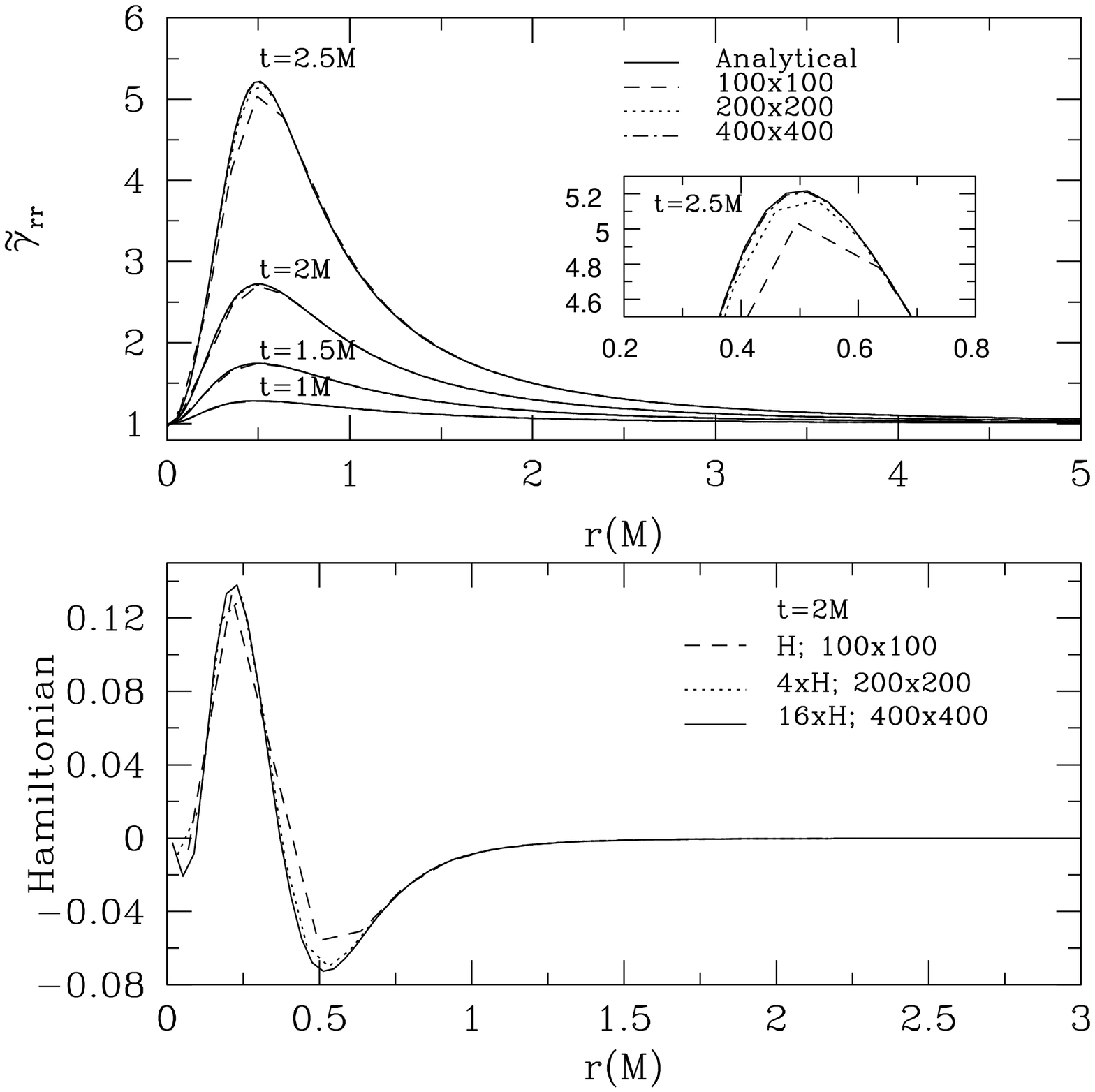}
\includegraphics[width=8.7cm, angle=0]{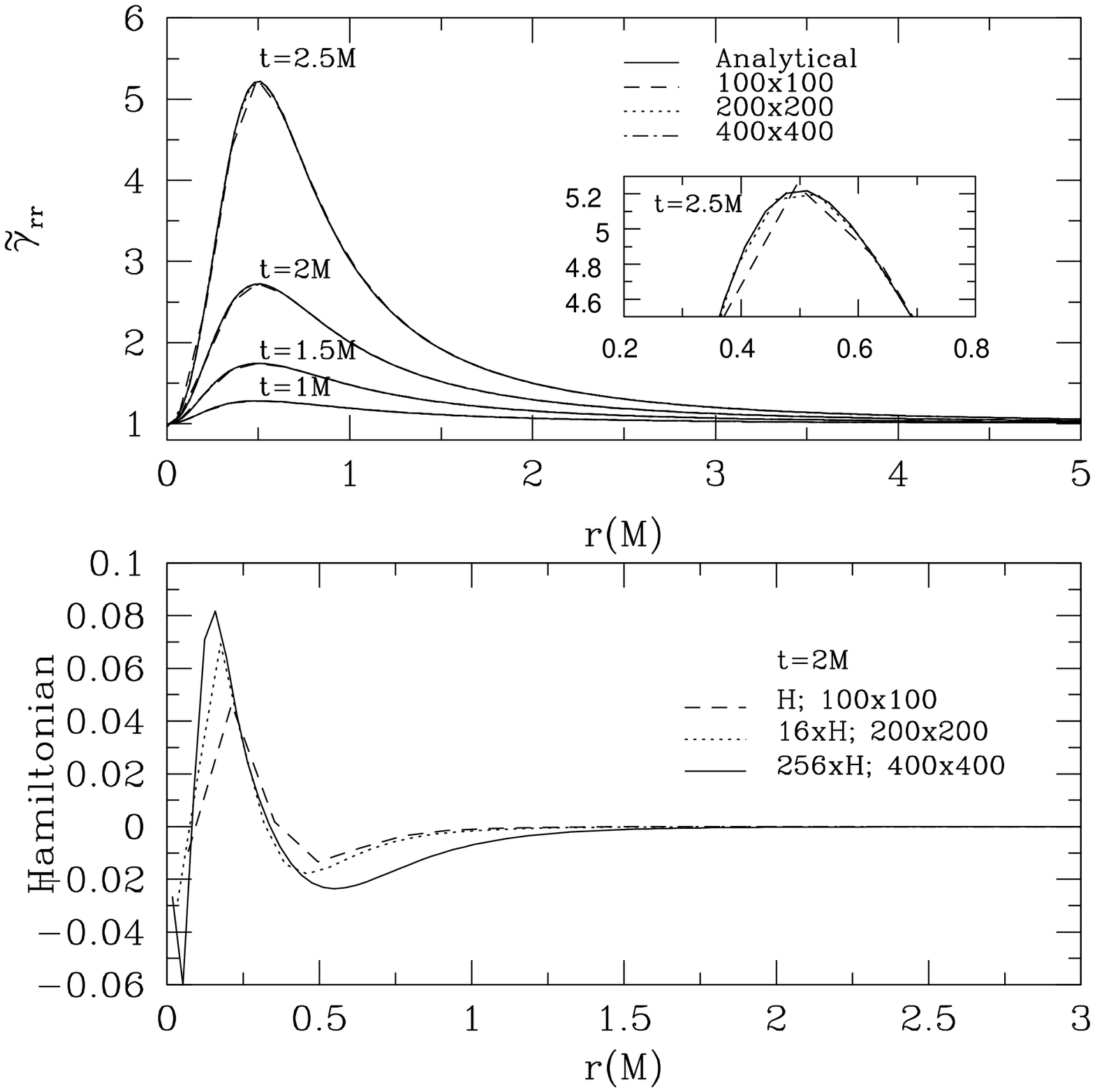}
\caption{ \label{fig1} Top-left panel: evolution of the conformal metric
  component $\tilde{\gamma}_{rr}$ along the diagonal of the
  grid. Different lines refer to the numerical solution with resolutions
 of 100, 200, and 400 grid points in each direction. The analytical
  solution is also shown. Bottom-left panel: Hamiltonian constraints along
  the diagonal at time $t=2M$ for three different resolutions showing the expected
  second-order convergence. The right panels show the same
  quantities and the expected convergence rate when using fourth-order finite differencing.}
\end{figure*}
\begin{figure}
\includegraphics[width=8.7cm, angle=0]{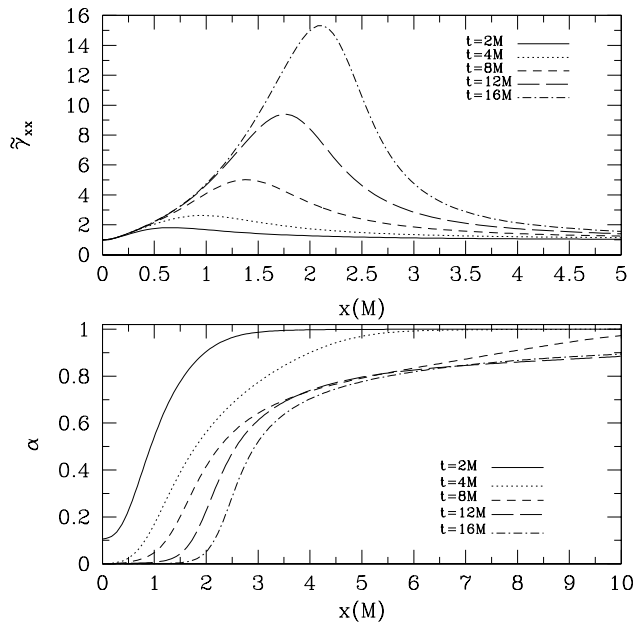}
\caption{ \label{fig2} Time evolution of the conformal metric
  component $\tilde{\gamma}_{xx}$ and the lapse $\alpha$ along the $x$-axis
  using $1+{\rm log}$ slicing.}
\end{figure}

\section{Vacuum tests: single BH evolutions }
\label{sec:vacuum-tests}

 The Schwarzschild metric in isotropic
 coordinates is used as initial data to test the ability of the code to
 evolve BH spacetimes within the fixed and moving puncture
 approaches. These initial data are such that the 3-metric is written in Cartesian
 coordinates as
\begin{equation}
\label{initdata}  
d\sigma^2=\psi^4(dx^2+dy^2+dz^2), 
\end{equation}
where the conformal factor is $\psi=(1 + {M}/{2r})$, $M$ being the mass of the BH. Here $r$ is the
isotropic radius, $r^2=x^2+y^2+z^2$. Thus, the spatial metric takes
the form $\gamma_{ij}=\psi^4 \tilde{\gamma}_{ij}$, where the
conformal metric is the flat metric. Initially the extrinsic curvature
is $K_{ij}=0$.

\subsection{Schwarzschild BH with geodesic slicing}

Following the fixed puncture approach, we first evolve these initial data with  geodesic slicing, that is setting $\alpha=1$ and $\beta^i=0$. Although, in these coordinates, the numerical evolution is known to last very short time and is expected to crash at $t=\pi M$ when the spacelike hypersurface reaches the physical singularity, there is an analytic solution for the evolution of the spacetime which can be used to compare to the numerical solution.

The top-left panel in Fig.~\ref{fig1} shows the evolution of the
radial metric component $\tilde{\gamma}_{rr}$, that is obtained from
the Cartesian metric functions, along the diagonal for different resolutions
together with the analytic solution at several time steps, performed
with second-order finite differences in space. The convergence of the
code can already be seen in this panel, but this is better appreciated
in the bottom-left panel, in which the Hamiltonian constraint is
plotted at $t=2M$ for the three different resolutions. Violations of
the Hamiltonian constraint are a measure of the numerical error, and
this figure shows that this error scales at the expected rate for
second-order convergence. Results obtained with a fourth-order finite
differencing are shown in the upper and lower right panels, where the
same quantities are plotted using this higher order
differencing. Clearly, using fourth-order finite differencing
increases the accuracy of the numerical evolutions, although the
convergence rate for the highest resolution is not exactly 
fourth-order in the whole computational domain. We note that in addition to the numerical error due to the
finite difference approximations for the spatial derivatives, there is
another source of numerical error due to the interpolation needed with the
Cartoon method.

\begin{figure}
\includegraphics[width=8.7cm]{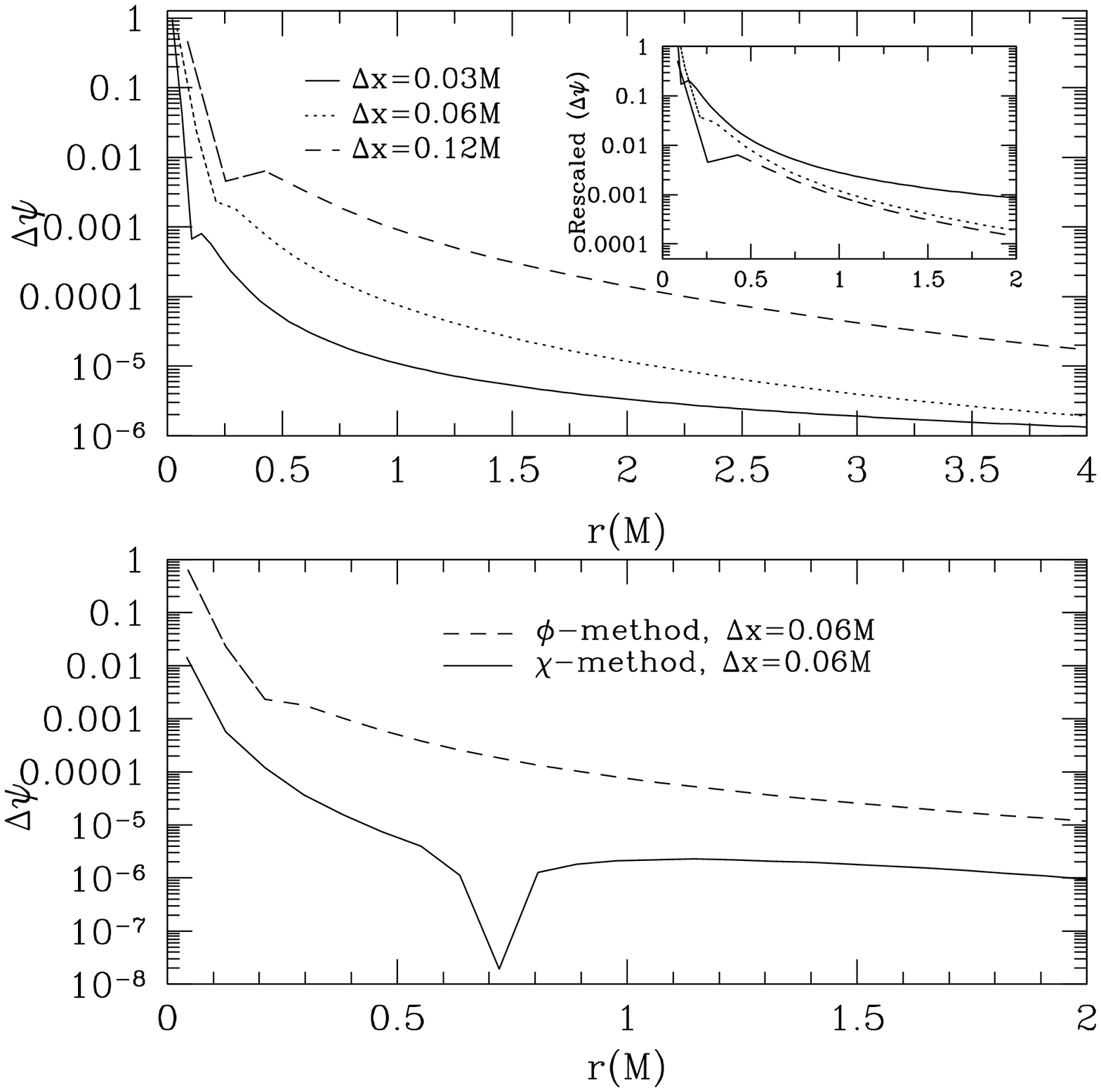}
\caption{In the top panel we consider as initial data the analytical
  solution proposed by Baumgarte and Naculich \cite{Baumgarte07} and show the deviations for the conformal
  factor from the exact solution for different resolutions using the $\phi$-method. In the inset, we show
  deviations rescaled to show the convergence. In the bottom panel, we show the deviations of the
  conformal factor using the same resolution, $\Delta x=0.06 M$,
  with the $\phi$-method as well as the $\chi$-method.}
\label{fig3} 
\end{figure}
\begin{figure}
\includegraphics[width=8.cm]{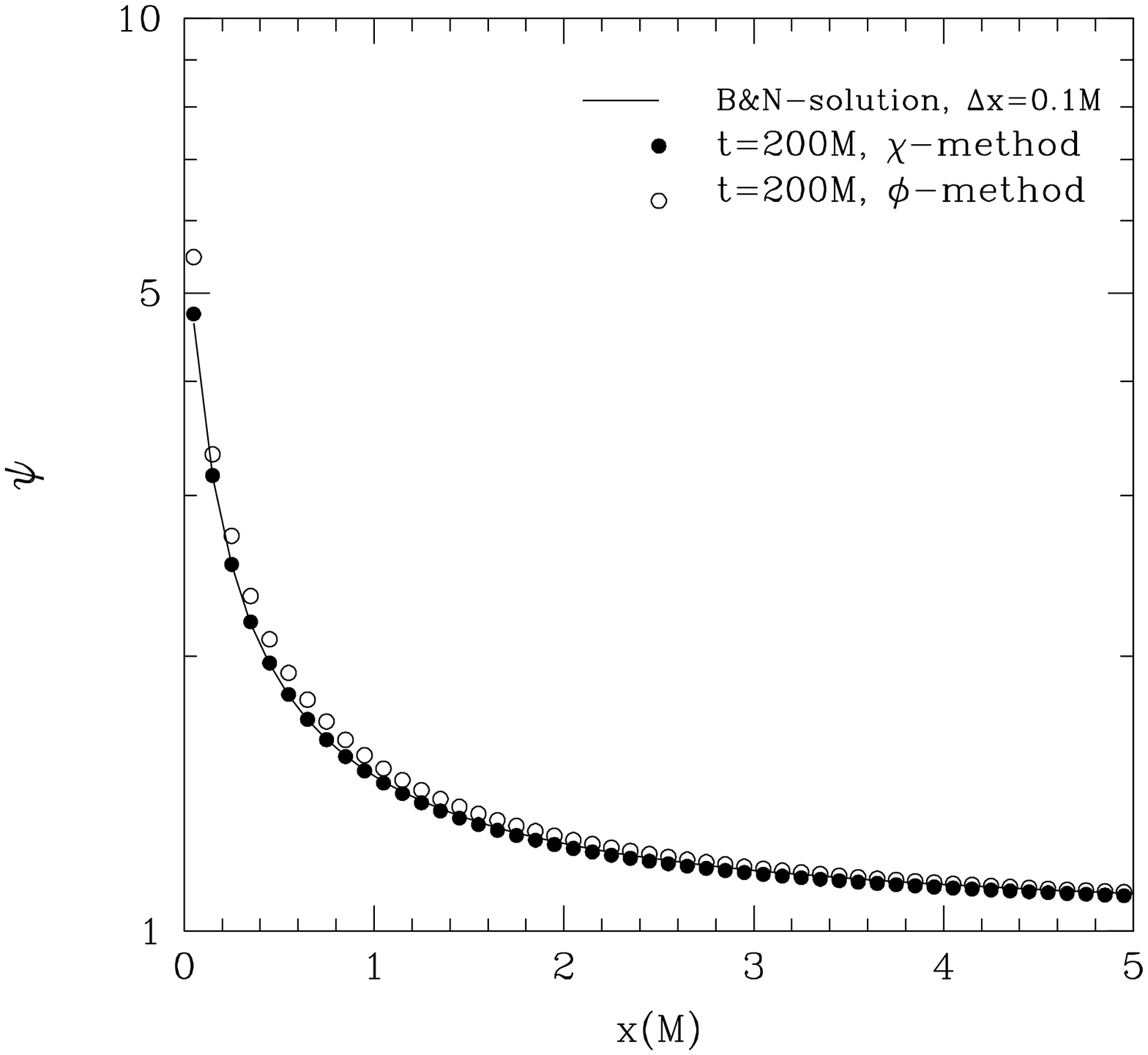}
\caption{We show, with a solid line, the conformal factor
  given by Baumgarte and Naculich \cite{Baumgarte07}, and with solid circles
  the numerical solution for the conformal factor at $t=200 M$ of Schwarzschild spacetime in
  isotropic coordinates initial data with the $\chi$-method and with empty
  circles the same quantity but computed using the $\phi$-method.}
\label{fig4} 
\end{figure}

\subsection{Schwarzschild BH with $1+\rm {log}$ slicing}

Next, we use a zero shift and a modified $1+\rm {log}$ slicing condition that permits the increase of the life of the simulation up to about 30--40$M$. The slicing condition used for this simulation is \cite{Imbiriba04},
\begin{equation}
\label{1+log3}  
\partial_{t}\alpha = -2 \alpha \psi^4 K. 
\end{equation}

Figure~\ref{fig2} displays the time evolution of one of the components
of the conformal three metric and the lapse function, and it proves
how the new gauge condition enables for longer simulations. For this
simulation we consider a grid resolution of $\Delta x=0.05M$ with 
$N_{x}\times N_{z}=200\times 200$ grid points. As expected, the metric function $\tilde{\gamma}_{xx}$ grows due to the grid stretching related to the collapse of the lapse. In order to increase further the length of the simulations of a single BH spacetime, we have implemented the moving puncture method, and its test is shown next.

\subsection{Schwarzschild BH and the moving puncture approach}

It has been pointed out \cite{Hannam07,Hannam07b,Bruegmann06}, that using the ``moving puncture method'' the numerical slices of a Schwarzchild BH spacetime reach a stationary state after about $40M$ of evolution.~\cite{Hannam07b} computed numerically this time-independent puncture data for the Schwarzschild spacetime  and showed that, as the evolution proceeds, the initial slice settles down to a maximal slice with $K=0$. This time-independent data has recently been obtained analytically by Baumgarte and Naculich \cite{Baumgarte07}.

With the aim of testing the ability of our code to perform long-term,
stable and accurate, evolutions of a single BH, we first check
the evolution of the BH puncture analytical data given  by
\cite{Baumgarte07}, with the gauge conditions given by
Eqs.~(\ref{1+log1})--(\ref{shift2}), and with two different
methods for the evolution of the conformal factor, the $\phi$-method,
which uses Eq.~(\ref{ecfactor}), and the $\chi$-method which, instead
uses the evolution Eq.~(\ref{chi}). For the damping parameter in the evolution equation
for the shift, we set $\eta=0.3/M$. 

In the upper panel of
Fig.~\ref{fig3}, we plot the absolute deviations for the conformal
factor between our numerical solution and the exact solution at
$t=9M$ obtained with the $\phi$-method, for three runs with different
resolutions ($\Delta x=0.12M,0.06M,0.03M$, with 
$N_{x}\times N_{z}=150\times 150,300\times 300,600\times 600$). Deviations for the
conformal factor are shown in the inner region of the grid, which 
at $t=9M$ is not affected by the outer boundary conditions. The inset,
shows these deviations from the exact solution, but rescaled assuming
fourth-order convergence. We note approximate fourth-order convergence
for the two lower resolutions runs ($\Delta x=0.12M,0.06M$), but,
although still converging, the convergence rate decreases
from fourth-order for the highest resolution run ($\Delta
x=0.03M$) particularly in the region outside the AH. For this resolution the error between the numerical and the
analytical solution in the region outside the AH is of the order of
$10^{-5}$--$10^{-6}$, and we find that increasing further the resolution does not
reduce significantly this error.

In the bottom panel of Fig.~\ref{fig3} we plot again the deviations for the
conformal factor, however these are now computed using the
$\chi$-method as well as the $\phi$-method, and obtained with the same
resolution, $\Delta x=0.06M$, at $t=9M$. Clearly, the
deviations from the exact solution obtained with the $\chi$-method are
smaller than with the $\phi$-method. In fact, these results are to be
expected, as $\phi$ diverges like $\rm {log}$ $r$ near the puncture,
while $\chi$ behaves like $r^{2}$ near the puncture.
   
We next perform a comparison between the time-independent analytical
data \cite{Baumgarte07} and the late time numerical solution of a
Schwarzschild spacetime time, in isotropic coordinates, choosing
initially $\alpha=\psi^{-2}$. For this comparison we use the same
gauge conditions as above, grid spacing $\Delta x=\Delta z=0.1M$ and a
grid with $N_{x}\times N_{z}=300\times 300$ points to cover the
computational domain. In Fig.~\ref{fig4}, we plot with a solid line
the time-independent analytical solution for the conformal factor,
while solid circles indicate the numerical solution for the conformal factor of
the Schwarzschild spacetime initial data at $t=200M$, computed using
the $\chi$-method and empty circles correspond to the same quantity but using the
$\phi$-method. Both simulations are compared at a sufficiently late time
to allow us to evaluate the effect of the outer boundary conditions on the late
time evolution. Clearly, the numerical solution agrees well with
the analytical solution, both for the $\chi$ and $\phi$ methods. 
However, again the $\chi$-method gives better results in the region
near the puncture. In addition, we observe that the outer boundary
conditions do not affect the long-term stability and accuracy of these
simulations.      

\section{Hydrodynamic tests}
\label{sec:hydro-tests}

\begin{figure*}
\includegraphics[width=8.7cm, angle=0]{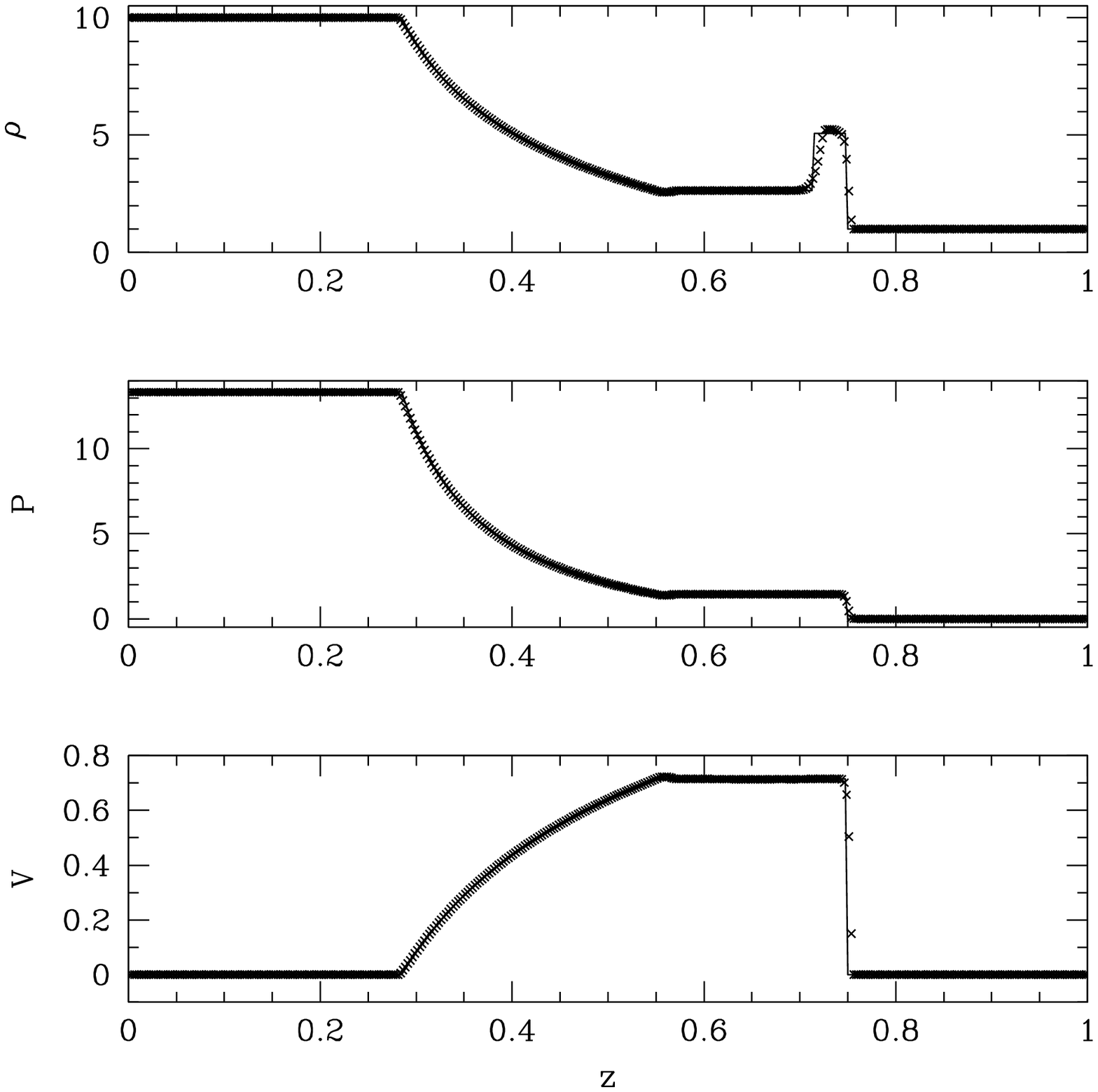}
\includegraphics[width=8.7cm, angle=0]{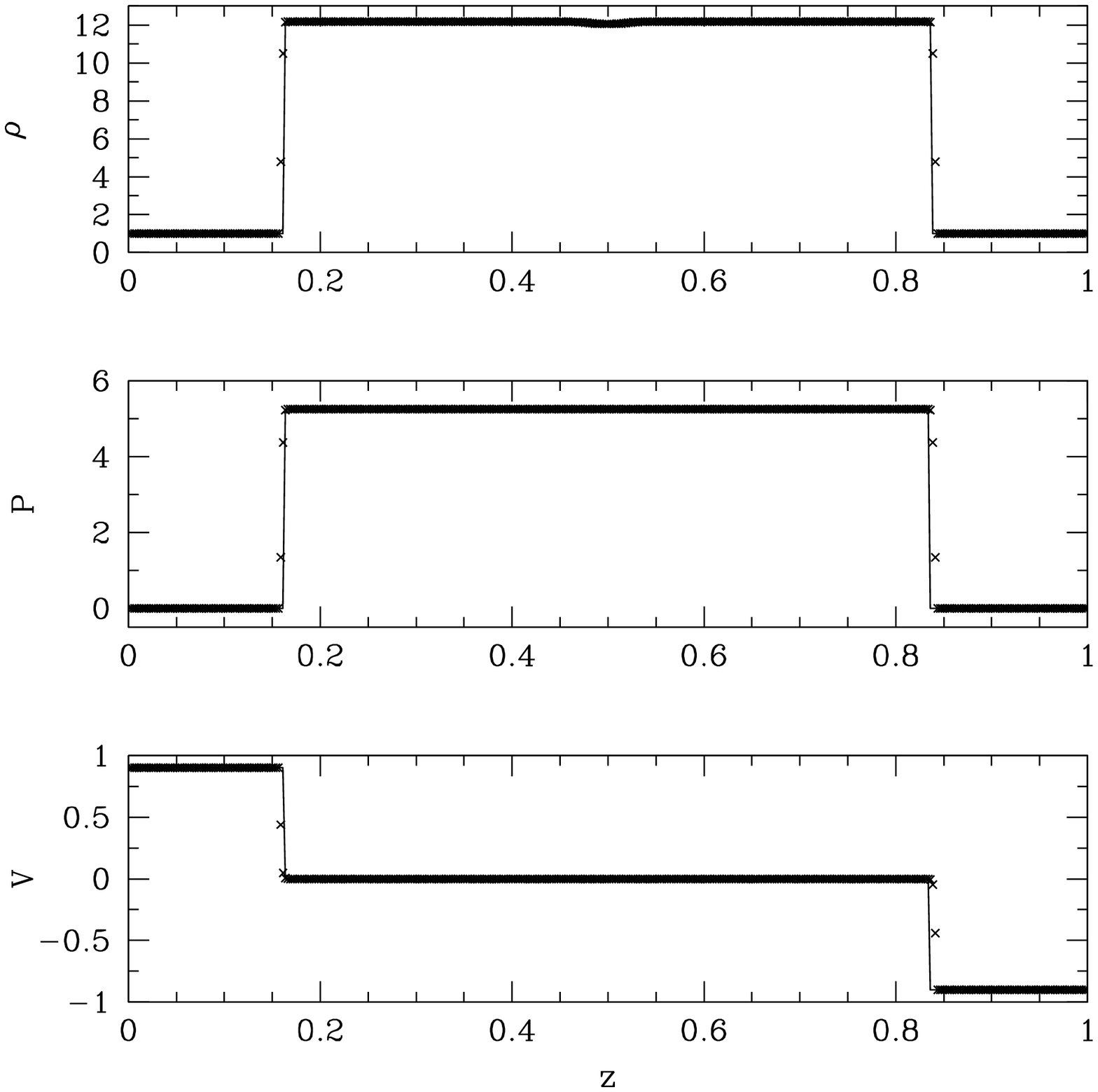}
\caption{Left panel: comparison of the numerical profiles of density, pressure and
velocity with the analytical solutions  for the solution of 
a one dimensional shock tube problem at time $t=0.3$. Right panel: wall shock
problem with $v=0.9$ at time $t=1.6$. The exact solution
is represented by the solid line and numerical solution by crosses.}
\label{fig5} 
\end{figure*}

\subsection{Relativistic shock tube test}

As a first test of the solution of the relativistic hydrodynamic equations we perform a standard one dimensional shock-tube problem (a Riemann problem) in flat spacetime. This is a common test to assess hydrodynamical codes (see e.g.~\cite{Marti96}). In this test, two uniform and different fluid states are initially separated by an interface which is then instantaneously removed. The initial data for the  left state is $P_{L}=13.33, \rho_{L}=10.00, v_{L}=0.00$ and for the right state $P_{R}=0.66\times 10^{-6}, \rho_{R}=1.00, v_{R}=0.00$, with the initial interface located at $z=0.5$. The fluid is assumed to be an ideal fluid with $\Gamma=5/3$.

The results of the numerical evolution are shown in the left panels of
Fig.~\ref{fig5}, where we plot the profiles at time $t=0.3$ of the pressure, density
and $z$-component of the fluid velocity for a shock-tube test along
the $z$-direction. The solid line represents the exact
solution of the shock-tube problem computed using the public domain
code {\tt RIEMANN} developed by J.M. Mart\'{\i} and
E. M\"{u}ller~\cite{Marti03}. The numerical solution is represented by
crosses. We use 400 grid points in the $z$-direction and a grid
spacing of $\Delta z= 1/400$. The particular combination of schemes
used for this hydrodynamical test comprise the Roe solver, MC
cell-reconstruction, and ICN for the time update. 

\subsection{Relativistic planar shock reflection test}

In our second test in Minkowski spacetime we carry out a relativistic
planar shock reflection problem along the $z$-direction with the following
initial conditions  $P=10^{-6}$, $\rho=1.0$ in the entire domain,
while $v=0.9$ for $z<0.5$, and $v=-0.9$ for $z>0.5$. We use an ideal
fluid equation of state with $\Gamma=4/3$. We show results of the
evolution in the right panels of Fig.~\ref{fig5}, where we plot the
profiles of the solution at time $t=1.6$ for the pressure, density and
$z$-component of the fluid velocity. Again, the solid lines refer to
the analytic solution while the numerical solution is represented by
crosses. The schemes used for this simulation are the HLLE solver,
minmod cell-reconstruction, and ICN for the time update. 

\subsection{Relativistic spherical shock reflection test}

The initial configuration  for this test problem consists of a medium
with uniform density ($\rho=1$) and pressure ($P=2.29 \times
10^{-5}(\Gamma -1)$) where $\Gamma=4/3$, with constant spherical inflow
velocity $v_{\rm in}=-0.9$  and Lorentz factor $W_{\rm in}$. Initially, at $t=0$,
the gas collides at the center of symmetry which forms a strong shock
wave that propagates upstream. The analytic solution for this test has the following form \cite{Marti97}

\begin{equation}
\rho(r,t) = \left \{ \begin{array} {cc}
	\left (1 +  \frac{|v_{\rm in}|t}{r} \right )^2, & r>v_{\rm s}t, \\
	\left (1+\frac{|v_{\rm in}|}{v_{\rm s}}\right )^2 \sigma, & r<v_{\rm s}t,
	\end{array} \right.
\end{equation}
with the compression ratio given by
\begin{equation}
\sigma = \frac{\Gamma+1}{\Gamma-1}+\frac{\Gamma}{\Gamma-1}(W_{\rm in}-1),
\end{equation}
and the shock velocity by
\begin{equation}
v_{\rm s} = \frac{\Gamma-1}{W_{\rm in}+1}W_{\rm in}|v_{\rm in}|.
\end{equation}

This test problem has been proposed by \cite{Aloy99} to test the
ability of a three-dimensional special relativistic hydrodynamics code
in Cartesian coordinates to keep the spherical symmetry of the
solution. For our axisymmetric hydrodynamics code in
Cartesian coordinates, this is a two-dimensional test problem, which
allow us to evaluate the solution of the hydrodynamic equations both in the $x$ and
$z$ directions. We use $N_{x}\times N_{z}=400\times 400$ grid points
with $\Delta x=\Delta z= 1/400$ to cover the computational domain. The schemes
used for this simulation are the HLLE solver, minmod
cell-reconstruction, and ICN for the time update.

\begin{figure}
\includegraphics[width=8.cm]{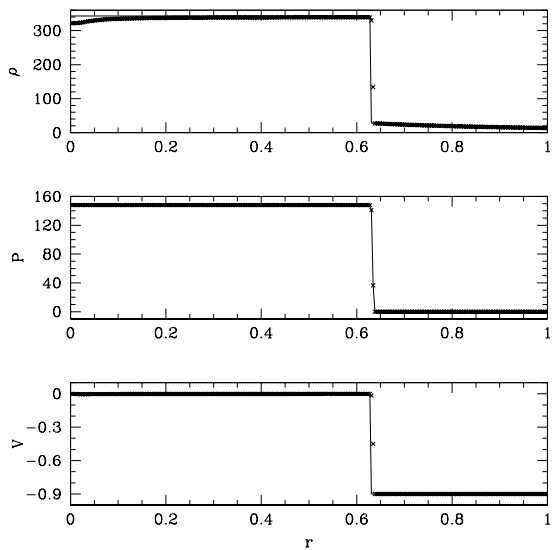}
\caption{Comparison of the numerical profiles of density, pressure and
velocity with the analytical solutions  for the solution of 
a relativistic spherical shock reflection test at time $t=3$}
\label{fig6} 
\end{figure}

 We show results of the evolution in Fig.~\ref{fig6}, where we plot
 the profiles of the solution at time
 $t=3$ for the density, pressure and fluid velocity along the
 diagonal. The solid line represents the analytic solution while the 
 numerical solution is represented by crosses.

To quantify the error we computed the relative global error between
the numerical and the analytical solution defined by $\epsilon_{\rm
rel}= \epsilon_{\rm abs}/[\sum_{i,k}|w(x_{i},z_{k},t_{n})|\Delta x
\Delta z]$, where the absolute global error is $\epsilon_{\rm
abs}= \sum_{i,k}|w^{n}_{i,k} - w(x_{i},z_{k},t_{n})|\Delta x \Delta
z$, $w^{n}_{i,k}$ is the numerical solution and $w(x_{i},z_{k},t_{n})$
is the analytical solution. We find that the relative global errors
for the density, pressure and velocity are $2.1\%$, $1.1\%$ and
$0.6\%$ respectively. As for the previous shock-tube test and planar shock
reflection test, the agreement achieved between the analytic and numerical solution is remarkable. 

\subsection{Spherical relativistic stars in a fixed spacetime}
%
\begin{figure*}
\includegraphics[angle=0,width=8.7cm]{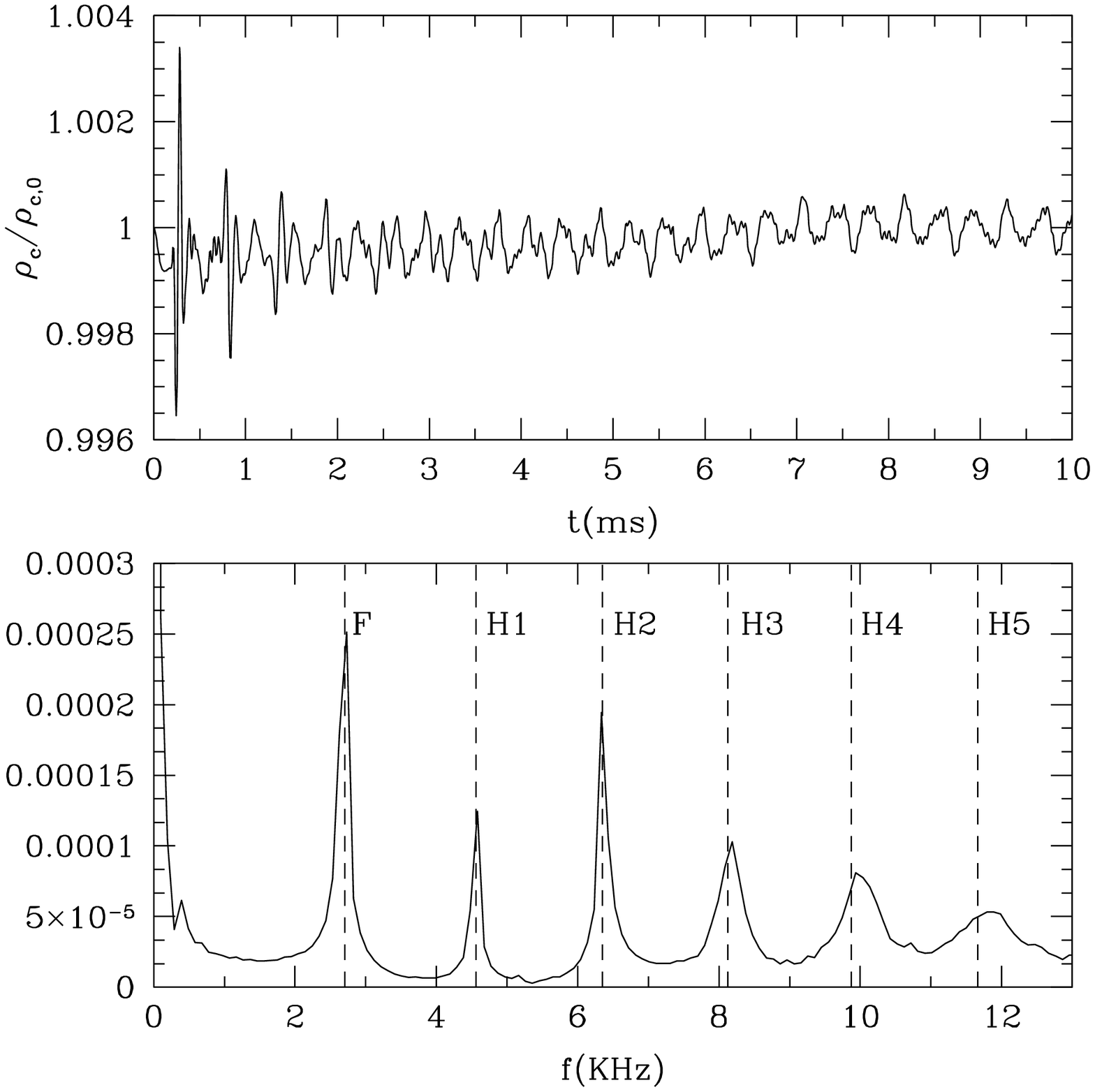} 
\includegraphics[angle=0,width=8.7cm]{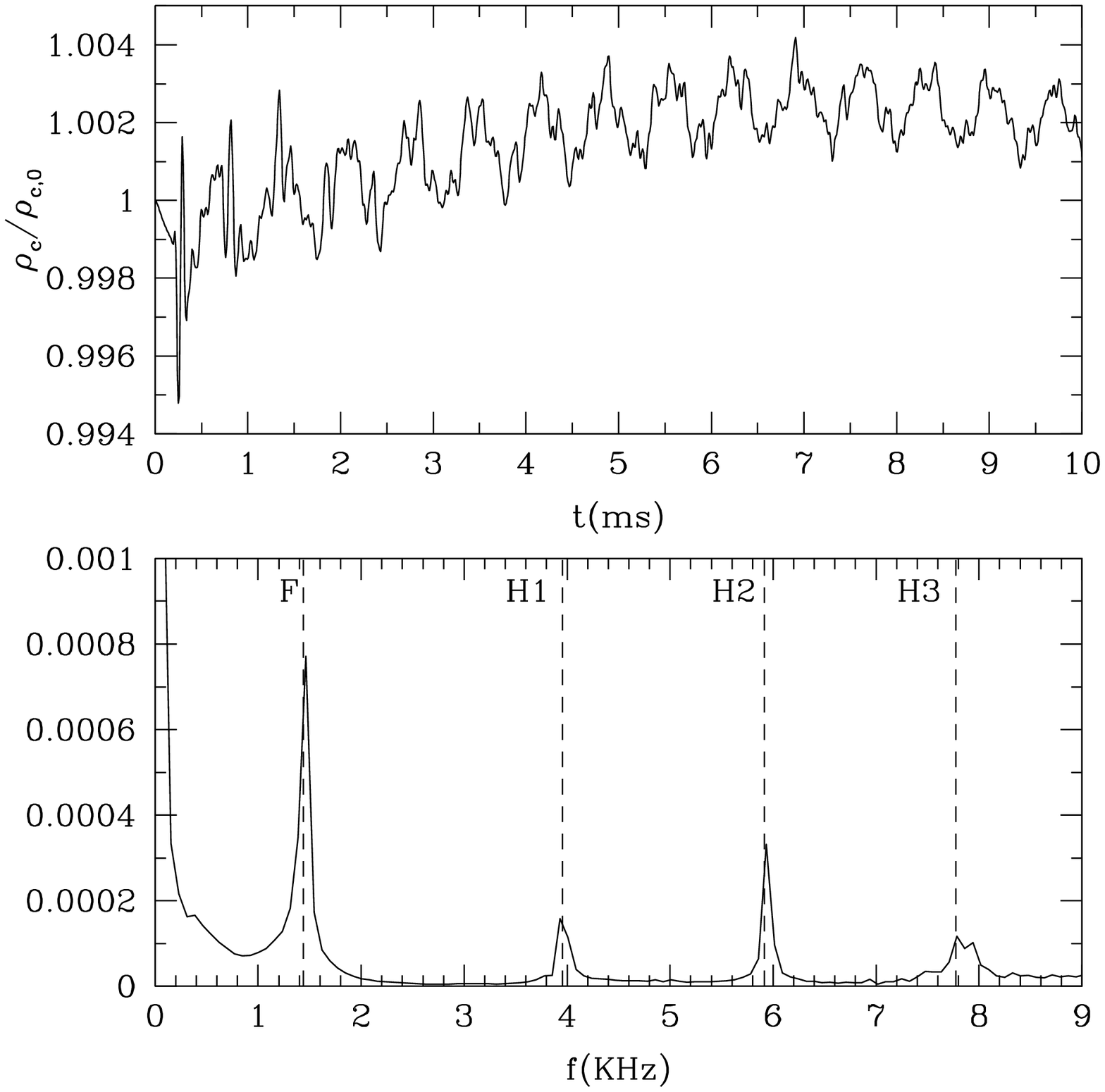}
\caption{Top panels show the time evolution of the normalized central
  density, in the Cowling approximation (left panel), and in a
  dynamical spacetime evolution (right panel). Power spectrum of the evolution of the central rest-mass density for an $M=1.4$, $\kappa=100$, $N=1$ polytrope in the Cowling approximation (bottom-left panel). F represents the frequency of the fundamental mode and H1-H6 are the first six overtones computed by \cite{Font00c}. Correspondingly, the bottom-right panel shows the same quantities in a dynamical spacetime evolution of the same TOV star. The frequencies computed by \cite{Font02c} are displayed with dashed vertical lines.} 
\label{fig7} 
\end{figure*}

For the next test we use the solution of the Tolman-Oppenheimer-Volkoff equations  to assess the capability of the code to perform long-term stable numerical simulation of a NS in equilibrium. In order to check the hydrodynamical evolution independently from the spacetime evolution we follow the common approach of keeping   the spacetime fixed during the numerical evolution. This is known as the Cowling approximation \cite{Cowling}, in analogy with the corresponding approximation in perturbative studies of stellar oscillations in which the metric components (or the gravitational potential in the Newtonian framework) are kept constant. 

It has been shown \cite{Font02c} that the truncation errors of the
finite difference representation of the PDEs are enough to excite
small periodic radial oscillations which manifest themselves  as
periodic variations of the hydrodynamical (and spacetime)
quantities. The power spectrum of the evolution of the central density
provides the frequencies of the fundamental mode of oscillation and of
its overtones, which can be compared with the corresponding
frequencies computed by  perturbative techniques. Although the excited
oscillations are purely numerical in origin ({\it i.e.} their
amplitude converges to zero as the resolution increases) they still
represent the oscillation modes of the relativistic star and their
frequencies should agree with the eigenfrequencies computed by linear
perturbation analysis.  For the purposes of further testing the code
and compare with independent results we focus on an initial TOV model
that has been extensively investigated numerically by
\cite{Font00c,Font02c}. This model is a relativistic star with $N=1$,
polytropic constant $\kappa=100$ and central rest-mass density
$\rho_{c}=1.28\times 10^{-3}$ so that its mass is $M=1.4$, its baryon
rest-mass $M_{*}=1.5$ and its radius $R=9.59$. We evolve these initial data with our non-linear code
and compute the power spectrum of the evolution of the central
rest-mass density. We note that in the simulations of spherical stars
in equilibrium presented below, 
we use the polytropic equation of state during the evolution.

In the upper-left panel of Fig.~\ref{fig7} we plot the time evolution of the central rest-mass density for a simulation with $\Delta x=0.15$. We observed that the truncation errors at this resolution are enough to excite small periodic radial oscillations, visible in this plot as  periodic variations of the central density. We see that the damping of the periodic oscillations of the central rest-mass density is very small during the whole evolution, which highlights the low numerical viscosity of the schemes implemented. By computing the Fourier transform of the time evolution of the central rest-mass density we obtain the power spectrum, which is shown with a solid line in the lower-left panel of the figure, while the dashed vertical lines indicate the fundamental frequency and the first five overtones computed by \cite{Font00c} with a non-linear hydrodynamics code using spherical polar coordinates. The agreement found for the fundamental frequency and overtones is very good, with the relative error between the fundamental frequencies being less than 1\%.

\subsection{Spherical relativistic stars in a dynamical spacetime}

For our first test of the coupling of the Einstein equations and the general relativistic hydrodynamic equations we again use the TOV  solution. In analogy to what happens in the Cowling approximation, truncation errors excite small oscillations in the star. Now, however, the truncation errors come not only from the hydrodynamic part of the code but also from the spacetime part solving the full set of Einstein equations.

We computed the eigenfrequencies of the coupled evolution of the same
TOV star with  $\kappa=100$, $N=1$ and central rest-mass density
$\rho_{c}=1.28\times 10^{-3}$ which has been discussed above under the
assumption of a fixed background spacetime. Again, the radial
oscillations excited by truncation errors manifest in the time
evolution of the central rest-mass density. The coupling to the
spacetime increases the amplitude of the oscillations, and also shifts
the frequencies of the modes towards lower frequencies. This is shown
in the lower-right panel of Fig.~\ref{fig7} which displays the power spectrum of the evolution of the central rest-mass density (solid line), and the eigenfrequencies computed by \cite{Font02c} (dashed lines). We note that the locations of the frequency peaks for the fundamental mode and the two overtones is in very good agreement, with the relative error in the fundamental frequencies being less than 1\%.

\begin{figure}
\includegraphics[angle=0,width=8.cm]{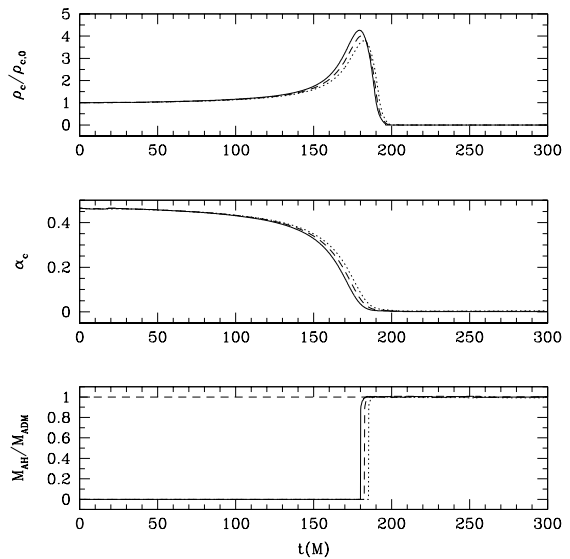} 
\caption{Time evolution of the normalized central density, the central
  lapse function, and mass of the AH in units of the ADM mass of the system
  for the collapse of a marginally stable spherical star to a
BH. Dotted lines represent results obtained with the lowest
resolution, dashed lines with the medium resolution while solid lines
with the highest resolution.}
\label{fig8} 
\end{figure}

\subsection{Gravitational collapse of marginally stable spherical relativistic stars}
We next test the capability of the code to
follow BH formation with the gravitational collapse to
a BH of a marginally stable spherical relativistic star. For this
test, we consider a $\kappa=100$, $N=1$ polytropic star with central
rest-mass density $\rho_{c}=3.15\times 10^{-3}$, its mass is $M=1.64$ and its baryon
rest-mass $M_{*}=1.79$. In order to induce the
collapse of the star we initially increase the rest-mass density by
0.5$\%$. 

We present numerical results for the simulation of the gravitational
collapse of a marginally stable spherical relativistic star performed
with $\Delta x= \Delta z= 0.1M, 0.075M, 0.05M$ to show the convergence
of the code. With these resolutions the star is
initially covered by approximately $60$, $80$ and $120$ points. The
hydrodynamics equations are solved using the Roe solver and MC
reconstruction, while the Einstein equations are solved with
fourth-order finite differencing and the gauge conditions given
in Eqs.~(\ref{1+log1})--(\ref{shift2}), with $\eta=0.5$. Time integration is done with
RK4. We plot in Fig.~\ref{fig8} the time evolution of the normalized
central density (top panel), of the lapse function
at the center (middle panel), and of the mass of
the AH (Eq.~\ref{Mah}) in units of the ADM mass of the system (bottom
panel). Dotted lines represent results obtained with the lowest
resolution, dashed lines with the medium resolution while solid lines
with the highest resolution. Overall, as the collapse proceeds the star increases its
compactness, and is reflected in the increase of the central density
as shown in the top panel. The middle panel
shows the characteristic collapse of the lapse function at the center
of the star indicating the formation of a BH. The most unambiguous signature of the
formation of a BH during the simulation is the formation of an
AH. Once an AH is found by the AH finder, we monitor the evolution of
the AH area, and also of its mass which is plotted, in the bottom
panel of Fig.~\ref{fig8}. This panel shows that the formation of the
BH delays with decreasing resolution, since the increase of the
central density slows down because numerical dissipation is larger for
the lower resolution runs. An AH is first found at approximately at
$t=180M$ ($t=0.88$ ms) for the simulation with $\Delta x= \Delta z=
0.05M$. This figure also shows that the mass of the AH relaxes to the ADM
mass of the system. The difference in the ADM mass and the mass of the
AH when we stop the evolution at $t=300M$, about $120M$ after
the AH is first found, is less than $1\%$.

This test shows that the code is able to follow BH formation and
its subsequent evolution for many timescales. We note that, unlike
the results discussed in~\cite{Baiotti06}, we do not need to add any numerical dissipation to the
evolution equations for the spacetime variables and gauge quantities
to perform this simulation. Instead, we rely only
on the gauge choice used here to follow the formation and evolution of the BH
formed as a result of the gravitational collapse.

\begin{figure}
\includegraphics[angle=0,width=8.cm]{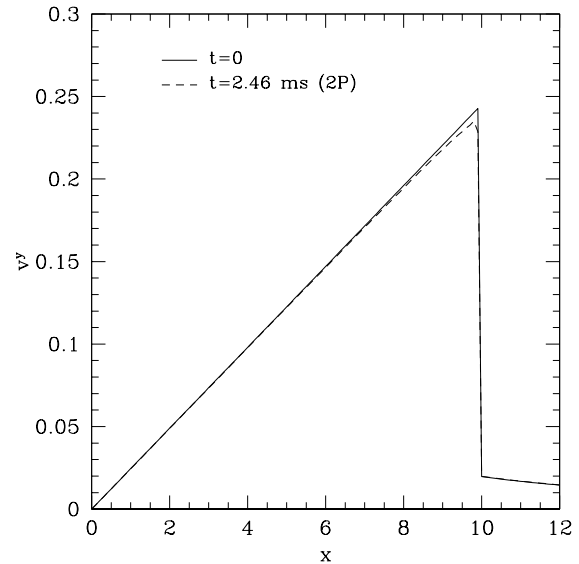} 
\caption{Profile of the $y$-component of the velocity of a rapidly
rotating relativistic star along the $x$-axis, at the initial time
(solid line) and at time $t=2.46$ ms, which corresponds to 2
rotational periods.}
\label{fig9} 
\end{figure}

\subsection{Rapidly rotating relativistic stars}
\begin{figure*}
\includegraphics[width=8.8cm,angle=0]{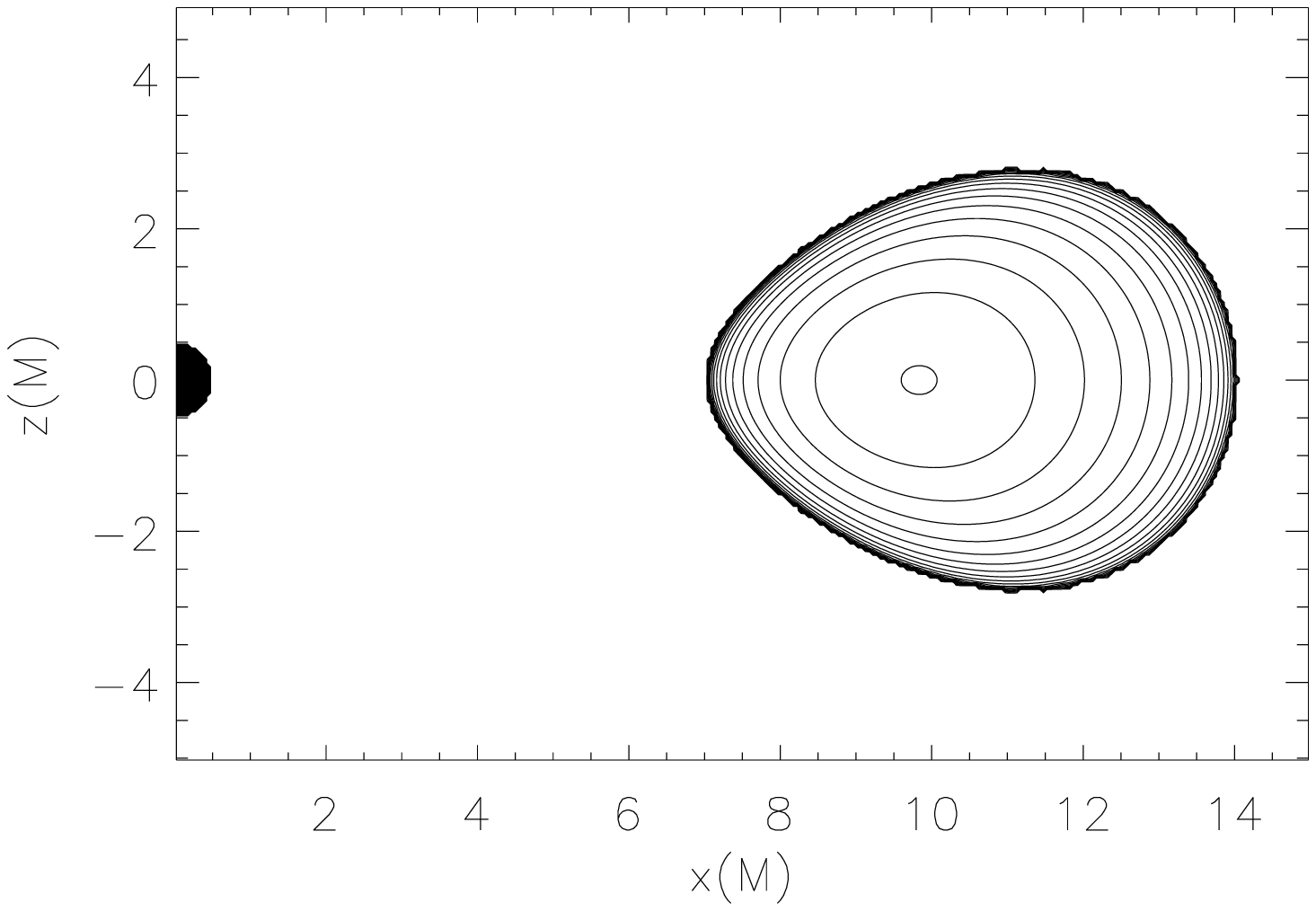}
\includegraphics[width=8.8cm,angle=0]{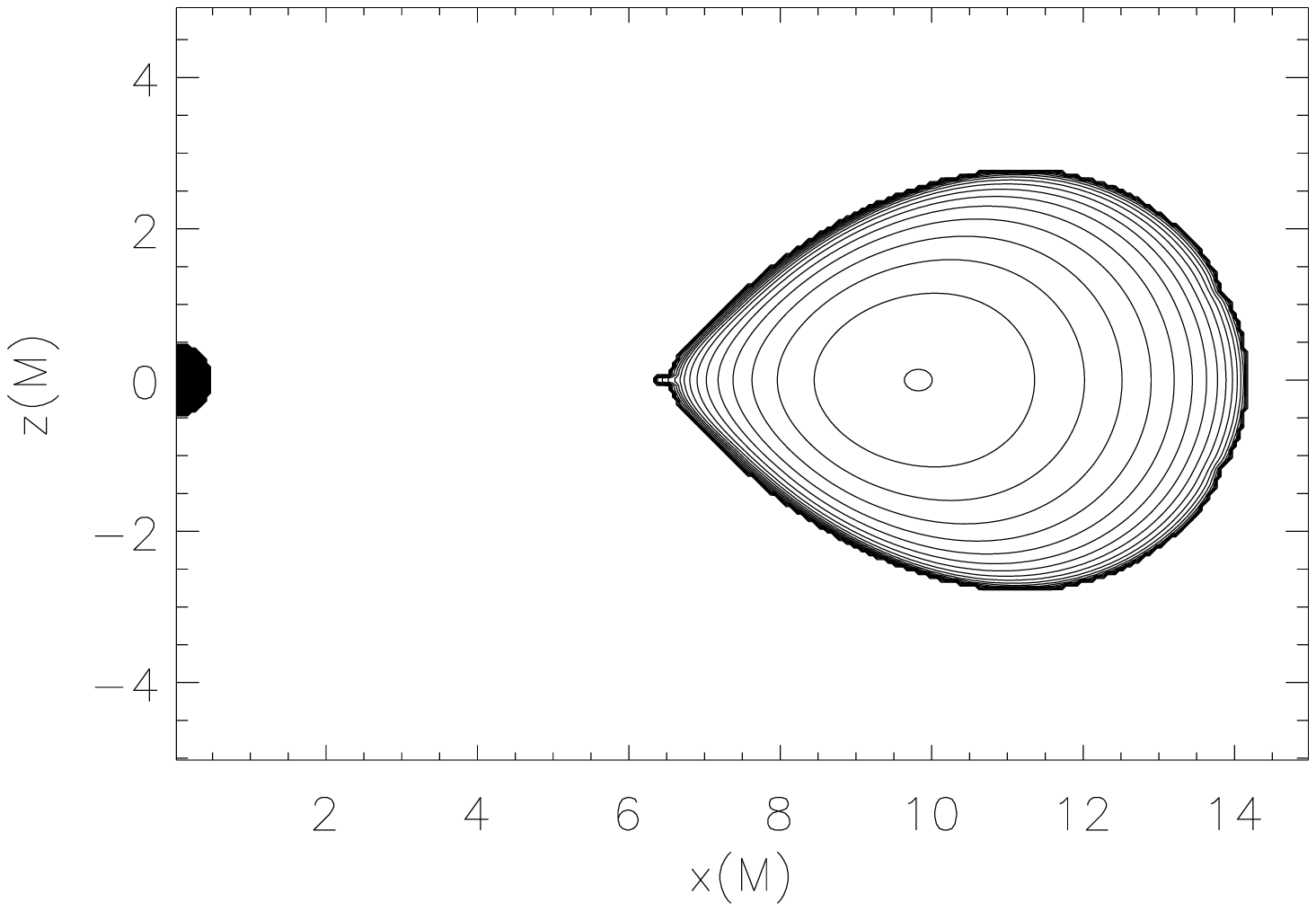}
\caption{Isocontours of the logarithm of the rest-mass density of the torus. The left panel shows the configuration at the
  initial time and the right panel the corresponding distribution
  after $5$ dynamical timescales (approximately $1000M$). The equilibrium solution is preserved
  to high accuracy.}
\label{fig10}
\end{figure*}
The evolution of stable rapidly rotating relativistic stars is a more
demanding test than all previous ones, as this involves testing parts of the code that are now used since there is a non-zero $y$-component of the shift vector. The initial data used for this test are the numerical solution of a stationary and axisymmetric equilibrium model of a rapidly and uniformly rotating relativistic star with angular velocity $\Omega$, which has been computed using the {\tt{Whisky}} code \cite{Baiotti03a,Baiotti06}. The initial data for this equilibrium rotating star are initially computed in spherical polar coordinates and then transformed to Cartesian coordinates using the standard coordinate transformation.

Here, we consider a uniformly rotating polytropic star with $\Gamma=2$,
$\rho_{c}=1.28\times 10^{-3}$, and rotating at $92\%$ of the allowed
mass-shedding limit for a star with the same central rest-mass density. The ratio of the polar to equatorial
coordinate radii for this model is $0.7$, its mass is $M=1.57$ and its baryon
rest-mass $M_{*}=1.69$. For this simulation we use the
{``1+log''} condition for the lapse, and the {``Gamma-freezing''}
condition for the shift vector, with $\eta=3.0$. We use the so-called
$\Gamma$-law equation of state during the evolution.

For this test, the outer boundary of the grid is placed about three
times the equatorial radius. We check that during the evolution, the
profiles of the different variables are kept very close to their
initial value for several rotational periods, which is shown in
Fig.~\ref{fig9}. Here we display the profile of the $y$-component of the velocity, $v^{y}$, along the $x$-axis at the initial time, $t=0$ (solid line) and at $t=2.46$ ms (dashed line), with the latter corresponding to two rotational periods. As has been shown by \cite{Stergioulas01,Font02c}, the small difference originating at the surface of the star after two rotational periods is expected since we used a second order reconstruction method with the MC limiter. Nevertheless, it shows that deviations of the numerical solution from the initial profile are very small.

\section{Simulations of self-gravitating torus in equilibrium around a BH}
\label{sec:torus}
%
\subsection{Initial data for BH and self-gravitating torus system }

The initial data for the numerical simulations of the system formed by
a BH and a self-gravitating torus in axisymmetric equilibrium have been recently computed by \cite{Shibata07}. The basic equations for the metric functions are derived assuming the 3+1 formalism, and the line element is written in the quasi-isotropic form as
\begin{equation}
\label{metricBH}
ds^{2} = - \alpha^{2}dt^{2} + \psi^{4}[e^{2q}(dr^{2}+r^{2}d\theta^{2})
+ r^{2} \sin^{2}\theta (\beta dt + d\varphi)^{2}].
\end{equation}
Where $e^{2q}$ denotes the conformal metric for the $rr$ and $\theta
\theta$ parts. The equations for the metric functions are solved in the BH puncture framework, and it is assumed that the puncture is located at the origin.

The equilibrium configurations for the matter are obtained by assuming
a perfect fluid stress-energy tensor, and adopting a polytropic
equation of state. The only nonzero components of the fluid
four-velocity are $u^{t}$ and $u^{\varphi}$, and initial
configurations can be constructed with either constant or non-constant
specific angular momentum distributions, defined as $j \equiv h
u_{\varphi}$. We refer to \cite{Shibata07} for a full explanation of
the construction of the initial model. For the simulations reported in
this paper,  we have considered a torus around a Schwarzschild BH (of
mass $M_{\mathrm BH}=1$). The adiabatic index is $\Gamma=4/3$ to mimic a degenerate  relativistic electron gas, and the polytropic constant $\kappa$ is fixed to $\kappa=0.0301262$ such that the torus-to-BH mass ratio, $M_{\mathrm t}/M_{\mathrm BH}$, is roughly 0.1. 

The initial model is computed on a spherical grid. Therefore, in order to use it as initial data for our evolution code, we transform into Cartesian coordinates and interpolate the values of the metric functions and hydrodynamic variables onto a cell-centered Cartesian uniform grid, retaining data for the $y=0$ plane.

 The chosen torus has a constant distribution of specific  angular momentum
 such that its inner and outer edges on the equatorial
 plane are located at $r_{\rm {in}}=7.1M$ and $r_{\rm {out}}=14.0M$, where $M$ is ADM mass of system.  Thus, such
 torus is initially covered with  approximately 140 points along the
 $x$-axis. The center of the torus is defined as the
 location at which the rest-mass density reaches its maximum and it is located at
 $r_{\rm {max}}=9.8M$. The dynamical timescale, which we choose as the
 orbital period at the center of the torus, is $t_{\rm orb}=223M$,
 that corresponds to about $1.1$ms for the case that $M=M_{\odot}$. 

 For these simulations we use an equally spaced uniform grid with a grid spacing $\Delta x=\Delta z=0.05M$ and a grid with $N_{x}\times N_{z}=600\times 600$ points to cover a computational domain, $0\leq x \leq L$ and $0 \leq z \leq L $, with $L=30M$. We use 4th-order finite differencing for the spacetime evolution, and Roe solver with MC reconstruction for the hydrodynamic evolution. The time integration is done with RK4.

\subsection{Dynamics in a fixed spacetime}
\label{sec:dynamics-fluid}

We have first investigated the equilibrium of the torus by performing numerical evolutions in a fixed spacetime, which nevertheless has the contribution coming from the self-gravity of the torus itself. This permits for one more test of the ability of the code to keep the stationarity of a fluid configuration initially in equilibrium. The accuracy check of the evolution is performed by comparing the stationarity and conservation of different local and global fluid quantities over a timescale which is several times the dynamical one. 

We show in Fig.~\ref{fig10} the isocontours of the logarithm of the rest-mass density of the torus as computed at the initial time $t=0$ (left panel) and at $t=1100M$ (right panel), the later corresponding to 5 dynamical timescales, when the code was stopped after approximately $3\times 10^4$ iterations, with no sign of the presence of numerical instabilities. These snapshots of the rest-mass density distribution clearly show the stationarity of the initial model for a sufficiently long period of time, and that the morphology of the torus remains unchanged for the duration of the simulation. In addition, they reflect that the  treatment of the atmosphere in the vacuum region does not perturb the equilibrium, and does not affect the dynamics during the evolution. 

\begin{figure}
\includegraphics[angle=0,width=8.7cm]{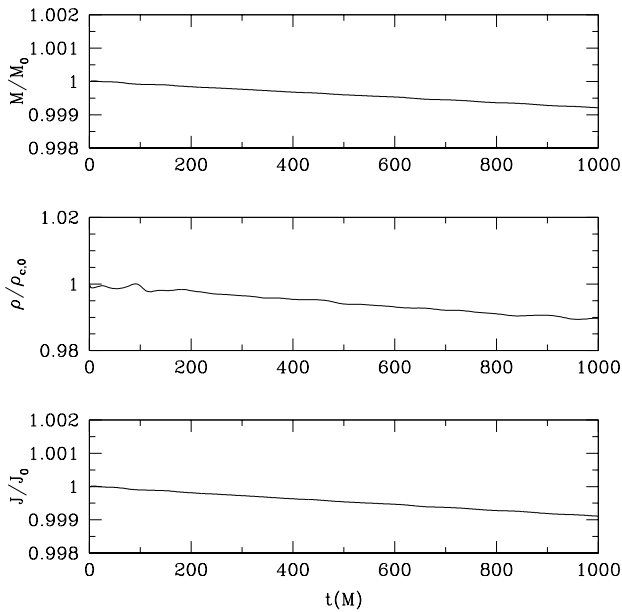} 
\caption{Time evolution of the total rest-mass, central
  rest-mass density and total angular momentum, each of them normalized to
  its initial value, for the evolution of a self-gravitating torus on a fixed spacetime.}
\label{fig11} 
\end{figure}

Other quantities that provide a more precise measure of the stationarity of the torus
are displayed in Fig.~\ref{fig11}, in which we plot the time evolution
of three fluid quantities normalized to their initial values. On the
top panel, we plot the time evolution of the total rest-mass of the
torus, normalized to its initial value. At the end of the simulation,
the difference between the final total rest-mass with respect to the
initial value is less than $10^{-3}$. This decrease of $0.1\%$ of the
total rest-mass is due to numerical error during the evolution, as the
rest-mass is not conserved exactly. However, this error does not affect significantly the stability of the torus for
the duration of our simulation. Similar results are obtained for the
time evolution of the other two quantities,  the central rest-mass
density and the total angular momentum, which  are displayed in the
lower two panels of Fig.~\ref{fig11}, respectively. The  central
rest-mass density, after a short initial transient phase, settles down
to a stationary value which only differs after $t=1000M$ by 1$\%$ from the initial one. This provides a strong evidence of the ability of the code to keep the torus in equilibrium for evolutions longer than the characteristic dynamical timescales of these objects. 
\begin{figure}
\includegraphics[angle=0,width=8.7cm]{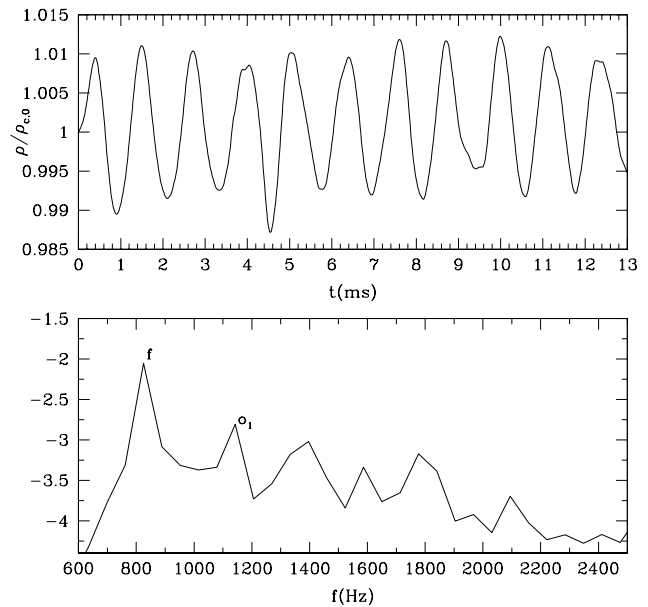} 
\caption{The upper panel shows the time evolution of the normalized central rest-mass density
    of a perturbed torus model, while the lower panel displays the power
    spectrum obtained after Fourier transforming the time evolution of
    the central rest-mass density. The fundamental mode and first
    overtone frequencies are denoted by $f$ and $o_{1}$. The mass of the BH is taken to be $1M_{\odot}$.}
\label{fig12} 
\end{figure}

 Notwithstanding the use of Cowling approximation, we can nevertheless extract some information about the oscillating behavior of these type of self-gravitating tori. We note that in a series of recent papers, \cite{Zanotti03,Rezzolla03b,Montero04,Zanotti05,Montero07}, it was shown that upon the introduction of perturbations, non self-gravitating relativistic tori in equilibrium manifest a long-term oscillatory behavior lasting for tens of orbital periods. An important feature of these axisymmetric $p$-mode oscillations of accretion tori is  that the lowest-order eigenfrequencies appear in the harmonic sequence $2:3$. Overall,  it was found that the $2:3$ harmonic sequence was present with a variance of $\sim 10\%$ for tori with a constant distribution of specific angular momentum and with a variance of $\sim 20\%$ for tori with a power-law distribution of specific angular momentum. The departure from the $2:3$ harmonic sequence  depends on a number of different elements that contribute to small deviations, such as the vertical size of the tori, the BH spin, the distribution of specific angular momentum, the EOS considered, and the presence of a small but nonzero mass-loss, which can all influence this departure.

In order to trigger the phase of small oscillations of the
self-gravitating torus in our simulations, we perturb the equilibrium
solution by adding a small perturbation in the $x$-component of the
velocity  (we recall that in equilibrium the $x$ and $z$-components of
the velocity are zero). The oscillations reflect in the time evolution
of the different fluid quantities, for instance, the central rest-mass
density, which  we Fourier-transform to obtain the resulting power
spectra.  This shows distinctive peaks at the frequencies that can be
identified with the quasi-normal modes of oscillation of the disk. In
Fig.~\ref{fig12}  we present, in the upper panel, the time evolution
of the normalized central rest-mass density of the perturbed
model. Due to the initial perturbation, the torus shows a persistent
oscillating phase around its equilibrium position. We follow the
evolution for about 13 dynamical timescales, and compute the power
spectrum obtained from the normalized central rest-mass density, which
is shown in the lower panel.  A rapid look at this figure reveals that
the torus power spectrum shows features which are very similar to
those mentioned above for non self-gravitating tori (see e.g. Fig.~4
of \cite{Zanotti05}). Namely, it can be
clearly identified in the spectrum a fundamental mode $f$, and a first
overtone $o_1$. Interestingly, the ratio of the fundamental mode and
the first overtone also show approximately the $2:3$ harmonic relation
with $o_1/f\sim 1.4$. In addition, the power spectrum reveals a series
of higher frequency peaks, which roughly coincide with linear
combinations of the $f$ mode and first overtones. However, simulations
lasting for longer timescales are needed to unambiguously identify
these peaks. 
 
\begin{figure}
\includegraphics[angle=0,width=8.9cm]{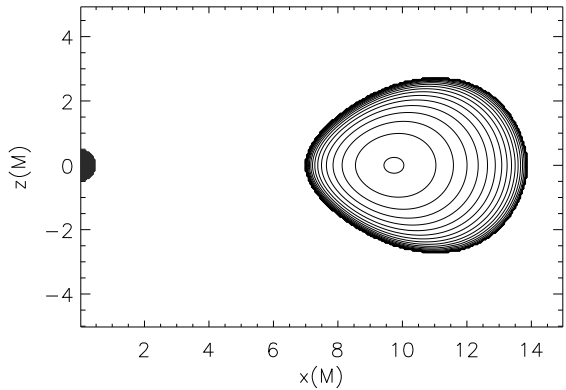} 
\caption{Isocontours of the logarithm of the rest-mass density of the
  torus during the evolution on a dynamical spacetime at
  $t=600M$.(Compare with the left panel of Fig.~\ref{fig10}.)}
\label{fig13}
\end{figure}

\begin{figure}
\includegraphics[angle=0,width=8.7cm]{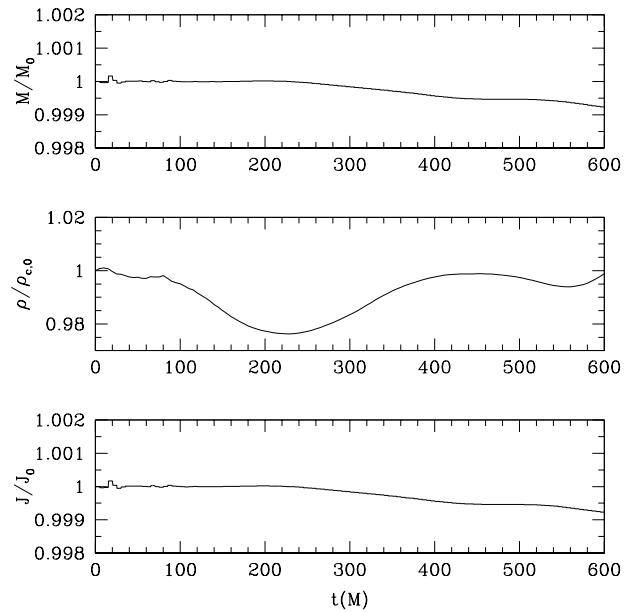} 
\caption{Time evolution of the total rest-mass density, central
  rest-mass density and total angular momentum, each of them normalized to
  its initial value, for the evolution on a dynamical spacetime.}
\label{fig14} 
\end{figure}

\subsection{Evolutions on a dynamical spacetime}
\label{sec:dynamics-full}

Next, we evolve the same torus and BH initial data in a fully dynamic
way, solving the Einstein equations coupled to the general relativistic hydrodynamics
equations. For such simulations we use the gauge conditions
given in Eqs.~(\ref{1+log1})--(\ref{shift2}) together with the
$\chi$-method, which allow for long-term and stable evolutions 
of the puncture.

 In our simulations of the system formed by a BH plus a self-gravitating torus,
 the spacetime evolution is highly dynamical until $t\simeq 30M$ due to the
 initial adjustment of the gauge, which produces a small pulse in the
 metric functions that propagates outwards. Outgoing radiative
 boundary conditions for the spacetime variables at the outer
 boundaries, permit this initial pulse to leave the grid. Despite
 this initial transient phase, the torus remains in equilibrium during the
 evolution, and its morphology is kept very close to the initial
 profile even several hundred $M$ beyond the first orbit when we stopped
 our simulation. In Fig.~\ref{fig13}, we plot the isocontours of
 the logarithm of the rest-mass density of the torus at time $t=600M$, 
 which is close to $3$ dynamical timescales. This figure clearly
 demonstrates that the configuration remains in equilibrium. Other
 quantities also exhibit this behavior. In particular, in
 Fig.~\ref{fig14}, we plot the time evolution of the total rest-mass of the
torus (top panel), the central rest-mass density (middle panel), and the total angular momentum
(lower panel), each normalized to its initial value, for the duration
of the simulation. At the end of the simulation,
the difference between the final total rest mass with respect to the
initial value is less than $0.1\%$. Although it is due to accumulated numerical error
during the evolution, this violation of the
rest mass does not affect significantly the stability of the torus for
the duration of our simulation. Similar results are obtained for the
time evolution of the other two quantities,  the central rest-mass
density and the total angular momentum. We note that, truncation
errors at the resolution employed seem to be enough to trigger small
oscillations of the torus around its equilibrium as shown in the
 evolution of the central rest-mass density.

 It is worth to mention, that in order to maintain the torus in equilibrium for
 several hundred $M$, we find crucial not to replace the
 $\tilde{\Gamma}^{i}$ by $-\partial_{j}\tilde{\gamma}^{ij}$ on the
 right hand side of the evolution Eq.~(\ref{eCCF}) whenever is not
 differentiated. Replacing $\tilde{\Gamma}^{i}$ by
 $-\partial_{j}\tilde{\gamma}^{ij}$ excites larger amplitude
 oscillations of the torus, which can be more than $10\%$ larger than
 in the case where $\tilde{\Gamma}^{i}$ is not replaced by its
 definition. We note also that, as observed for single puncture simulations in vacuum,
 numerical errors in the spacetime evolution, especially near the puncture,
 are smaller with the $\chi$-method than with the $\phi$-method. We
 find that an additional ingredient that helps to maintain the torus equilibrium 
 for the duration of the simulations is the use of the $\chi$-method
 for the evolution of the conformal factor (we refer to
 Fig.~\ref{fig3} and Fig.~\ref{fig4} for a comparison of the
 $\chi$-method and  $\phi$-method). We find numerical indications that, at least in conjunction with the
 Cartoon method, these two procedures help to decrease the error in the violation of
 the Hamiltonian constraint.

\section{Conclusion}
\label{sec:conclusion}

In this paper  we  have presented a  new two-dimensional numerical code designed to solve the full Einstein equations coupled to the general relativistic hydrodynamics equations. The code is mainly intended for studies of self-gravitating accretion disks around BHs, although it is also suitable for regular spacetimes. 

Concerning technical aspects, the Einstein equations are formulated
and solved in the code using a reformulation of the standard 3+1 (ADM)
system, the so-called BSSN approach, in conjunction with the Cartoon
method to impose the axisymmetry condition under Cartesian
coordinates, and the puncture/moving puncture approach to carry out
BH evolutions. We note that a key and novel feature of the
code is that combines a fourth-order finite difference
approximation for derivative terms in the spacetime evolution with the
Cartoon method. Correspondingly, the general relativistic
hydrodynamics equations are written in flux-conservative form and solved with high-resolution, shock-capturing schemes. 

We have performed and discussed a number of tests to assess the
accuracy and expected convergence of the code, namely (single) BH
evolutions, shock tubes, and evolutions of both spherical and
rotating relativistic stars. We have also presented a simulation of the
gravitational collapse to a BH of a marginally stable spherical
star. We have shown that the code is able to handle the formation of a BH,
and to follow the BH evolution until we the simulation is stopped after
several hundred $M$. We remark that we did not add any numerical dissipation to the
evolution equations for the spacetime variables and gauge quantities
to perform this simulation; and exclusively relied 
on the gauge choice to follow the BH formation and its long-term
evolution. Overall, the code has passed those tests with
remarkable accuracy. 

 In addition, paving the way for specific applications of the code, we
 have also presented results from fully general relativistic numerical
 simulations of a system formed by a BH surrounded by a
 self-gravitating torus in equilibrium. First, simulations using a
 fixed spacetime evolution, have shown that the torus remains in
 equilibrium around its initial configuration for more than $5$
 dynamical timescales, when the simulation was stopped. Furthermore,
 after adding a small perturbation on the torus in equilibrium, we
 have followed the evolution of the torus through a phase of small oscillations for more than 13 dynamical timescales. The computation of the frequencies of the fundamental mode of oscillation and the first overtone revealed that these two frequencies appear close to the 2:3 harmonic relation observed for the case of non self-gravitating relativistic tori.

 Our simulation of the same model in a dynamical spacetime, showed
 that the code is able to keep the torus in equilibrium for the several hundred $M$
 that lasted the simulation. We have 
 found numerically, that using the $\chi$-method and not to replace the
 $\tilde{\Gamma}^{i}$ by $-\partial_{j}\tilde{\gamma}^{ij}$ on the
 right hand side of the evolution for the $\Gamma^{i}$ are two
 important issues that reduce the numerical error in the spacetime
 evolution, and therefore, help to maintain the torus in
 equilibrium. 

 Overall, the simulations performed indicate that the code is able to perform such
 simulations accurately and  for the sufficient duration needed to
 produce scientific results for one of the specific applications
 for which it has been designed, the investigation of the runaway
 instability of thick accretion disks around BHs. We point out that the last simulation presented in the
 paper is the first 2D axisymmetric simulation of the system formed
 by self-gravitating torus around a BH in a dynamical spacetime that has been
 ever carried out. In a follow up paper, we aim to present results from an
 expanded range of initial models, and to investigate in detail the
 dynamics of such systems, focusing on the development on the runaway
 instability, and the evolution of the BH due to the transfer of mass 
 and angular momentum from the torus.


\acknowledgments 

It is a pleasure to thank K. Uryu for many useful discussions,
as well as to L. Rezzolla and J.C. Miller for 
many valuable comments, and J.M. {Ib\'a\~nez} also for carefully
reading this manuscript. We also thank L. Baiotti for
computing the initial data for the rotating relativistic star
discussed in Section VF, computed with the {\tt RNSID} thorn by N. Stergioulas. P.M. has benefited from a VESF fellowship of the
European Gravitational Observatory (EGO-DIR-126-2005). This research
has been partially supported by the Spanish Ministerio de Educaci\'{o}n y
Ciencia (grant AYA2007-67626-C03-01). M.S. acknowledges
support from Japanese Monbukagaku-sho Grant Nos. 19540263 and
20041004. The computations have been
performed on the computer ``CERCA2'' of the Department of Astronomy
and Astrophysics of the University of Valencia. 


\bibliographystyle{apsrev}


\end{document}